\documentclass[11pt, preprint]{elsarticle}
\usepackage{amsmath,amssymb,amsthm}
\usepackage{bm}
\usepackage{mathtools}
\usepackage{mathrsfs}
\usepackage{graphicx}
\usepackage{float}
\usepackage{subcaption}
\usepackage{multirow}
\usepackage{tabularx}
\usepackage{booktabs}
\usepackage{siunitx}
\usepackage{enumitem}
\usepackage{xcolor}
\usepackage[margin=1in]{geometry}
\usepackage{url}
\usepackage{hyperref}
\usepackage{cleveref}
\hypersetup{colorlinks=true, citecolor=blue, linkcolor=blue, urlcolor=blue}

\journal{Engineering Fracture Mechanics}

\newcommand{\disp}{\boldsymbol{u}}
\newcommand{\dmg}{d}
\newcommand{\xvec}{\boldsymbol{x}}
\newcommand{\dd}[1]{\mathrm{d}#1}
\newcommand{\td}{\text{d}}
\newcommand{\Vector}[1]{\boldsymbol{#1}}
\newcommand{\bfsigma}{\boldsymbol{\sigma}}
\newcommand{\bfepsilon}{\boldsymbol{\epsilon}}
\newcommand{\bforce}{\boldsymbol{b}^\ast}
\newcommand{\sforce}{\boldsymbol{t}^\ast}

\newcommand{\nd}{--}

\usepackage{amsmath,amssymb,amsthm}
\begin{document}

\begin{frontmatter}

\title{Fracture-Bench: A Controlled Benchmark Suite for Machine Learning Surrogates in Phase-Field Fracture Mechanics}

\author[jhu,dsai]{Meenu Krishnan U.}
\author[jhu,dsai]{Maryam Hakimzadeh}
\author[jhu,dsai]{Indrashish Saha}
\author[jhu,dsai]{Meer Mehran Rashid}
\author[jhu,dsai]{Lori Graham-Brady}
\author[jhu,dsai]{Somdatta Goswami\corref{cor1}}
\cortext[cor1]{Corresponding author}
\address[jhu]{Department of Civil and Systems Engineering, Johns Hopkins University, Baltimore, MD 21218, USA}
\address[dsai]{Data Science and AI Institute, Johns Hopkins University, Baltimore, MD 21218, USA}

\begin{abstract}
Machine-learning (ML) surrogates are increasingly proposed as replacements for phase-field fracture simulations in tasks that require many solutions, such as uncertainty quantification, inverse analysis and design. There is, however, little controlled evidence on which architectures can be trusted, for which fracture problems, and how far beyond their training data. We introduce Fracture-Bench, an open benchmark of two phase-field datasets released with fixed splits, common test metrics and a description of the physical regime each dataset covers. Dataset~1 contains 1,000 \texttt{AT2} simulations of multiple cracks growing from a notched hole in a hyperelastic plate under random, non-uniform biaxial boundary displacements (44 load steps, $128\times128$ grid). Dataset~2 contains 4,000 \texttt{PF-CZM} simulations of a pre-cracked plate with two inclusions that are each ten times stiffer or ten times more compliant than the matrix ($128\times256$ grid). Five architectures, a Fourier neural operator (FNO), DeepONet, Latent DeepONet, a UNet and a conditional diffusion model, are trained on identical splits and scored with one test metric per dataset, together with training time, inference time and model size. On Dataset~1, FNO and Latent DeepONet reach mean test errors of order $10^{-6}$, about an order of magnitude below UNet and DeepONet. On Dataset~2, the diffusion model is the most accurate, with an error about six times lower than that of UNet, but it needs \SI{706}{s} per prediction against \SI{6}{ms} for UNet. For every architecture, most of the error arises in the final load steps, once cracks localize, and Dataset~2 contains initial cracks less than one grid cell apart that end in qualitatively different crack paths. Architecture rankings therefore depend on the task, and errors averaged over densely sampled inputs say little about near-bifurcation cases. The data, codes, splits and regime descriptors are public, together with generalization splits and a contribution protocol for extending the benchmark to under-represented regimes such as mixed-mode, three-dimensional and experimentally paired fracture.
\end{abstract}

\begin{highlights}
\item Open benchmark of 5,000 phase-field fracture simulations with fixed splits
\item Five surrogates compared on accuracy, training cost, inference cost and size
\item No architecture is best on both datasets; rankings depend on the task
\item Errors concentrate where cracks localize and near crack-path bifurcations
\item Regime descriptors and generalization splits show where surrogates are tested
\end{highlights}


\end{frontmatter}

\section{Introduction}
\label{sec:introduction}

Fracture often sets the safe life of engineering components such as aircraft engines, pipelines, reactor pressure vessels and additively manufactured parts. It often occurs abruptly, propagates rapidly and can lead to catastrophic failure. Because fracture cannot yet be predicted with confidence, design and certification still rely on extensive testing and conservative margins. A national study for the US National Bureau of Standards estimated the annual cost of fracture in the United States at \$88 billion in 1978 dollars, about 4\% of the gross national product, of which roughly a third could have been avoided with the technology then available \cite{reed1983economic}. Fully qualifying a single additively manufactured material by empirical testing often requires many thousands of individual tests, costing millions of dollars and taking 5 to 15 years \cite{nist_am_qualification}. The NASA-sponsored Vision 2040 roadmap envisions product certification that relies heavily on simulation, with physical testing only where necessary \cite{liu2018vision2040}. Reaching that goal for fracture requires predictions that are both fast and trustworthy.

Griffith \cite{griffith1921} and Irwin \cite{Irwin1958} founded fracture mechanics on the energy release rate and the stress intensity factor. These criteria can be evaluated analytically only for simple geometries and loadings, and they do not describe the nucleation, branching or coalescence of cracks, which require solving the balance of momentum coupled with an evolving crack. Heterogeneous materials add difficulty, because spatial variations of stiffness and toughness make the crack path sensitive to the material layout \cite{krishnan2024adaptive}. Sharp-crack methods struggle with these features. The extended finite element method (X-FEM) \cite{moes_finite_1999} decouples the crack from the mesh through enrichment \cite{Khoei2014, Laborde2005, andresgonzalez-estradaRoleEnrichmentStatical2012}, but it needs crack tracking, which becomes cumbersome when cracks branch or merge, and its predictions depend on the propagation criterion and the mesh \cite{wu_computational_2019, jinErrorcontrolledAdaptiveExtended2017}. Cohesive zone models confine the crack to element boundaries. Phase-field models \cite{francfort1998revisiting, Bourdin2000, amor2009regularized, miehe2010thermodynamically, ambati2015review} and their cohesive extensions \cite{wuUnifiedPhasefieldTheory2017, Geelen2019} instead smear the crack over a band whose width is set by a length scale $\ell_o$, so that nucleation, branching and merging follow from energy minimization without tracking. The price is a fine mesh in the band and an iterative coupled solve at every load step. Adaptive sub-stepping \cite{guptaAutoadaptiveSubsteppingAlgorithm2020}, robust monolithic solvers \cite{bharaliRobustMonolithicSolver2022} and adaptive mesh refinement \cite{Heister2015, AREIAS2016322, goswamiAdaptivePhaseField2019, goswamiAdaptiveFourthorderPhase2020, hirshikeshAdaptivePhaseField2021, krishnan2022adaptive, guptaAdaptiveMeshRefinement} reduce this cost. Even so, a specimen-scale simulation can take hours, which puts Monte Carlo uncertainty propagation, Bayesian inference and design optimization out of reach. Surrogate models address this bottleneck by learning an inexpensive approximation from precomputed solutions.

Neural networks were first applied to fracture by \citet{Theocaris1993}. Later surrogates estimated failure paths and damage evolution from high-fidelity simulations \cite{Mudunuru2019}, but they predicted low-dimensional quantities and did not enforce the governing equations. Physics-informed neural networks (PINNs) build the governing equations or the energy functional into the loss, as \citet{goswami2020b} did for phase-field crack paths. A PINN, however, must be retrained for each new geometry, loading or material. Neural operators instead learn a mapping between function spaces, so one model covers a family of inputs and, in principle, any discretization. The most widely used are DeepONet \cite{lu2019deeponet} and the Fourier neural operator (FNO) \cite{li2020fourier}. Latent DeepONet \cite{kontolati2023learning} learns the operator in the latent space of an autoencoder, which helps for high-dimensional outputs. Physics-informed variants \cite{goswami2023physics} include a DeepONet trained with a variational loss to predict crack paths in quasi-brittle materials \cite{goswami2022physics}. Convolutional and generative models treat prediction as an image-to-image task on a fixed grid. The UNet \cite{ronneberger2015u}, an encoder--decoder with skip connections, has been applied to elasto-plastic fields in two-phase composites \cite{saha2024prediction} and to spall failure in ductile polycrystals \cite{saha2026numerical}. Denoising diffusion models \cite{ho2020denoising} generate the field by iterative denoising. They preserve sharp features that regression tends to smooth and return samples from a distribution, but they need many network evaluations per prediction. Trained surrogates can be orders of magnitude faster than the solvers they replace \cite{Rashid2022, rashid2023revealing}, including for phase-field simulations \cite{bonneville2025accelerating, hamdi2025towards}.

\begin{table}[!t]
\small
\centering
\renewcommand\arraystretch{1.15}
\caption{Regime descriptors of the two datasets. ``Effective input dimension'' counts the independent random variables that control the sampled inputs; for Dataset~1, it counts the GRF coefficients $\xi_k,\eta_k$ of the modes that carry 99.9\% of the variance of $f$. $h$ is the spacing of the benchmark grid. Dataset~1 is non-dimensional: its parameters approximately correspond to PMMA, all quantities are normalized by the shear modulus $\mu$, and $\lambda = 0$ (equivalently $\nu = 0$) is chosen to remove the effect of Poisson's ratio \citep{T1XFF19O_2025}.}
\label{tab:regime}
\begin{tabularx}{\textwidth}{>{\raggedright\arraybackslash}p{3.6cm}>{\raggedright\arraybackslash}X>{\raggedright\arraybackslash}X}
\toprule
 & \textbf{Dataset~1} & \textbf{Dataset~2} \\
\midrule
Dimension, kinematics & 2D, finite deformation & 2D plane strain, small strain \\
Domain & $2.2 \times 2.2$ (non-dimensional) & $4 \times 8$~mm \\
Bulk model & Neo-Hookean & Isotropic linear elastic \\
Material parameters & $\mu = 1$, $\lambda = 0$ ($\nu = 0$), $G_c = 1.721\times10^{-5}$, $\ell_o = 0.015$ & Matrix: $E = 377$~GPa, $\nu = 0.3$, $G_c = 10$~N/m, $\ell_o = 0.04$~mm; inclusions: $E$ and $G_c$ multiplied by 10 (stiffer) or 0.1 (less stiff) \\
Fracture model & \texttt{AT2}, vector phase field $\mathbf{d}\in\mathbb{R}^2$, crack-face contact & \texttt{PF-CZM}, scalar phase field $d$, linear softening \\
Loading & Displacement control, non-uniform biaxial tension & Displacement control, uniaxial tension (Mode~I) \\
Mode mixity & Nominally opening; local mixity at notches not quantified & Nominally Mode~I; local mixity at inclusions \\
Rate & Quasi-static, 44 increments & Quasi-static, 31 increments \\
Heterogeneity & Homogeneous; central hole with eight notches & Two circular inclusions of fixed size and position \\
Sampled inputs & GRF boundary profile $f$ & Crack-tip location $(x_c,y_c)$; inclusion materials \\
Effective input dimension & $\approx 4$ ($\xi_k,\eta_k$ for $k\le2$) & 2 continuous + 2 binary \\
Symmetry & Mirror symmetry about $x_1=x_2$ & None \\
Records (train/test) & 1,000 (900/100) & 4,000 (3,600/400) \\
Grid, spacing $h$ & $128^2$, $h \approx 0.0173$ & $128\times256$, $h \approx 0.0315$~mm \\
Crack band on grid & $\ell_o/h = 0.87$; $d>0.5$ band $\approx1.2h$ & $\pi\ell_o/h \approx 4.0$; $d>0.5$ band $\approx1.3h$ \\
Stored fields & $\mathbf{u}$, $\mathbf{d}$ at all 44 steps & $\mathbf{u}$, $d$, $E$ at all 31 steps \\
Known data issues & One training sample (index 257) stops after step 42; steps 43--44 are zero & 34 samples have no initial crack in $d(\mathbf{x},0)$ \\
Paired experiments & None & None \\
Not covered & 3D; dynamics; plasticity; cyclic loading; shear-dominated loading; geometry variation; experiments & 3D; dynamics; plasticity; cyclic loading; variation of inclusion size or position; experiments \\
\bottomrule
\end{tabularx}
\end{table}

Experience in other fields shows that such methods mature once an open, curated dataset exists. AlphaFold \cite{jumper2021alphafold} was trained on the Protein Data Bank, which began in 1971 with seven structures \cite{berman2000pdb}. Weather models trained on the ERA5 reanalysis \cite{hersbach2020era5} now have surpassed physics-based forecasts \cite{lam2023graphcast, bi2023pangu}, and have run operationally at the European Centre for Medium-Range Weather Forecasts since 2025 \cite{lang2024aifs, ecmwf2025aifs_operational}. Fracture has no comparable resource. Curated databases are licensed, industrial test data are proprietary, and open simulation data occupy a narrow regime. The largest collections are all two-dimensional and tensile: 400,000 dynamic phase-field simulations under uniaxial or biaxial tension \cite{hill2025fracturedata}, and 70,000 and 6,000 quasi-static brittle simulations in Mechanical MNIST Crack Path \cite{mohammadzadeh2022crackpath} and PFM-Fracture \cite{hamdi2025towards}. In our audit of eleven public sources, more than 99.7\% of roughly 700,000 simulations describe two-dimensional brittle specimens under quasi-static tension, and none covers mixed-mode loading, physically grown three-dimensional cracks or a matching experiment. Experiments share this bias: standard toughness tests measure the opening mode \cite{astm_e1820_25}, whereas many service failures involve mixed-mode cracks \cite{qian1996mixed}. Coverage matters because purely data-driven neural operators are usually reliable only within their training data and can fail under extrapolation \cite{zhu2023reliable, benitez2024ood}. Fracture adds sensitivity to inputs. In the first Sandia Fracture Challenge, thirteen teams made blind predictions of ductile tearing \cite{boyce2014sfc1}. Nominally identical specimens failed along tensile- or shear-dominated paths, a difference later traced to variation in the location of a hole, with out-of-tolerance specimens favoring the shear path \cite{kramer2019sfcfare}. No team had accounted for such tolerances \cite{kramer2019sfc3preface}. A fracture benchmark should therefore report not only an average error but also where in the input space the predictions hold.

Each new architecture, proposed by a research group, is usually demonstrated on its authors' own problem, the literature does not tell a practitioner which surrogate to choose for a given fracture problem. Differences in datasets, splits, preprocessing, losses, training budgets and metrics often outweigh the differences between the models themselves. Controlled comparisons exist for smooth problems, for example of DeepONet and FNO on common FAIR data \cite{lu2022comprehensive}, but general scientific-ML benchmarks such as PDEBench \cite{takamoto2022pdebench} and The Well \cite{ohana2024well} do not include fracture. For fracture, \citet{hamdi2025towards} introduced PFM-Fracture: 6,000 two-dimensional quasi-static brittle simulations spanning three energy decompositions, two loading conditions and 1,000 random initial cracks. They evaluated a PINN, an FNO and a UNet, with ensembling. Fracture-Bench complements that effort in four ways. First, it covers different physics: Dataset~1 combines finite-deformation hyperelasticity, a vector-valued phase field and random non-uniform boundary loading, and Dataset~2 treats cohesive fracture in a plate with inclusions ten times stiffer or ten times more compliant than the matrix. Second, it compares five architectures from three families (neural operators, convolutional networks and diffusion models) on accuracy, training and inference cost, and model size, with the datasets, splits and metrics fixed for all. Third, each dataset carries a regime descriptor of the governing variables it covers, the effective dimension of its inputs and what it lacks, and the benchmark defines splits that test generalization beyond random held-out samples. Fourth, the datasets, simulation codes, training code and an interactive dashboard are public, and \Cref{sec:extending} describes how to add architectures and fracture problems.

Both datasets lie within the dominant regime of open fracture data: they are two-dimensional and quasi-static. We do not claim that they fill the coverage gap described above. Their purpose is to provide a controlled starting point on which architectures can be compared on equal terms, and a format, including the regime descriptor and the contribution protocol, through which datasets from under-represented regimes can be added and compared on the same footing. Our contributions are:
\begin{itemize}[leftmargin=*,nosep]
    \item \textbf{Two high-fidelity simulation datasets} generated with published phase-field solvers \cite{hakimzadeh2022phase, krishnan2024adaptive}:
    \begin{itemize}
        \item \textit{Dataset~1:} 1,000 simulations of brittle fracture in a perforated hyperelastic plate under boundary loading drawn from a periodic Gaussian random field (GRF) ($128\times128$ grid, 44 load steps, \texttt{AT2} formulation) \cite{T1XFF19O_2025}; \url{https://doi.org/10.7281/T1XFF19O}.
        \item \textit{Dataset~2:} 4,000 simulations of cohesive fracture and crack--inclusion interaction in heterogeneous plates containing two circular inclusions, generated with adaptive mesh refinement ($128\times256$ grid, 31 load steps, \texttt{PF-CZM} formulation) \cite{T1RZFI3I_2025}; \url{https://doi.org/10.7281/T1RZFI3I}.
    \end{itemize}
    \item \textbf{A regime descriptor for each dataset} (\Cref{tab:regime}), which records the governing variables covered, the effective dimension of the sampled inputs, the resolution of the crack band on the benchmark grid and the information that the dataset does not contain.
    \item \textbf{A reference study} of five surrogate architectures, FNO, DeepONet, Latent DeepONet, UNet and a conditional denoising diffusion probabilistic model (cDDPM), trained on identical data splits and scored with common test metrics, together with training and inference wall-clock time, parameter count and temporal interpolation at load steps withheld from training.
    \item \textbf{An open, extensible platform.} The datasets, the simulation codes, the training and evaluation code (\url{https://github.com/Centrum-IntelliPhysics/fracture_bench}) and an interactive dashboard (\url{https://fracture-bench.github.io}) are public, and the benchmark is designed so that the community can add both new architectures and new fracture problems (\Cref{sec:extending}).
\end{itemize}

\noindent \Cref{sec:methodology} describes the two datasets, their regime descriptors, the surrogate architectures, the training and evaluation protocols, and how the benchmark can be used and extended. \Cref{sec:results} presents the benchmark results and discusses what they do and do not show. \Cref{sec:conclusion} summarizes the findings.

\section{Methodology}
\label{sec:methodology}

This section describes the datasets, the surrogate architectures and the evaluation protocols. Dataset~1 captures crack pattern evolution in a perforated hyperelastic domain under spatially varying boundary displacements using the \texttt{AT2} phase field
formulation%
, while 
Dataset~2 examines crack--inclusion interactions in a heterogeneous plate
solved with the \texttt{PF-CZM} phase field formulation.
For each dataset we state the problem setup, the governing equations and the key simulation parameters. Five surrogate architectures are then described, and finally the training and evaluation protocols that are applied to all of them.

\subsection{Phase-field fracture: governing equations}

Consider a linear elastic body $\Omega \subset \mathbb{R}^{2}$ with boundary $\partial\Omega$ partitioned into a Dirichlet
part $\partial\Omega_u$, on which displacements
$\boldsymbol{u}^{*}(\boldsymbol{x})$ are prescribed, and a Neumann part
$\partial\Omega_t$, on which tractions are applied, such that
$\partial\Omega_u \cup \partial\Omega_t = \partial\Omega$ and
$\partial\Omega_u \cap \partial\Omega_t = \varnothing$. Let
$\disp(\xvec)$ denote the displacement field and $\dmg(\xvec) \in [0,1]$
the scalar damage (phase-field) variable at a point $\xvec \in \Omega$,
with $\dmg = 0$ representing intact material and $\dmg = 1$ a fully
broken state. This scalar, small-strain formulation applies directly to Dataset~2 and is presented here as the common reference. Dataset~1 uses the vector-valued extension of \citet{hakimzadeh2022phase}, in which the phase field is a vector $\mathbf{d}$ whose magnitude $\|\mathbf{d}\| \in [0,1]$ plays the role of the scalar damage $d$ and whose direction gives the orientation of the crack normal, and in which the bulk response is neo-Hookean at finite deformation; the choices specific to each dataset are given in \Cref{sec:problem_statement}. Within the second-order phase-field
framework~\cite{goswamiAdaptivePhaseField2019}, the system Lagrangian
$\mathscr{L}(\disp, d)$ is written as
\begin{align} \label{eq:lagrangian}
\mathscr{L}(\disp, d) =
\mathcal{K}
- \int_{\varOmega}\psi_s(\bfepsilon, d)\,\mathrm{d}V
- \int_{\varOmega}\dfrac{G_{\mathrm{f}}}{c_{\alpha}}
  \left[\dfrac{\alpha(d)}{\ell_o} + \ell_o\,(\nabla d \cdot \nabla d)\right]
  \mathrm{d}V
+ \mathscr{P}(\boldsymbol{u}),
\end{align}
where $\bfepsilon = \tfrac{1}{2}\bigl(\nabla\disp +
\nabla\disp^{\mathsf{T}}\bigr)$ is the infinitesimal strain tensor,
$\psi_s(\bfepsilon, d)$ is the strain energy density of the damaged
solid, $G_{\mathrm{f}}$ is the material fracture energy (the energy
dissipated per unit area of crack surface), and $\ell_o$ is the
phase-field regularization length scale that controls the width of the
diffuse crack band. The geometric crack function $\alpha(d)$ governs the
spatial profile of the damage field across the crack band, and
$c_\alpha = 4\int_0^1 \sqrt{\alpha(\beta)}\,\dd{\beta}$ is a
normalization constant ensuring that the regularized surface energy
converges to $G_{\mathrm{f}}$ times the crack area in the limit
$\ell_o \to 0$. The kinetic energy $\mathcal{K}$ and the external work
$\mathscr{P}$ are given by
\begin{align} \label{eq:KE}
\mathcal{K} := \frac{1}{2}\int_{\varOmega}\rho\,\dot{\disp}\cdot\dot{\disp}
\,\mathrm{d}V,
\qquad
\mathscr{P} := \int_{\varOmega}\bforce\cdot\disp\,\td{V}
             + \int_{\partial\varOmega_{t}}\sforce\cdot\disp\,\td{A},
\end{align}
where $\rho$ is the mass density, $\dot{\disp}$ is the velocity field,
$\bforce$ is the prescribed body force per unit volume, $\sforce$ is the
prescribed surface traction on $\partial\varOmega_t$, and $\mathrm{d}V$
and $\mathrm{d}A$ denote volume and surface measures, respectively.

To prevent crack growth under compression and interpenetration of crack
faces, the strain energy density $\psi_s(\bfepsilon, d)$ in
\Cref{eq:lagrangian} is decomposed into tensile and compressive
contributions as
\begin{align} \label{eq:psi-split}
\psi_s(\bfepsilon, d) = \psi_0^{-}(\bfepsilon) + \omega(d)\,\psi_0^{+}(\bfepsilon),
\end{align}
where $\psi_0^{+}$ and $\psi_0^{-}$ are the tensile and compressive parts
of the undamaged elastic strain energy density, obtained here via the
principal (spectral) strain
decomposition~\cite{miehe2010thermodynamically}, and
$\omega(d)$ is the energetic degradation function, a monotonically
decreasing function with $\omega(0) = 1$ and $\omega(1) = 0$ that reduces
the tensile stiffness as damage accumulates. Only the tensile part is
degraded; the compressive part $\psi_0^{-}$ remains unaffected.

Applying Hamilton's principle of least action and enforcing the
irreversibility and boundedness of the damage field\footnote{Based on
Griffith's theory of fracture: the crack is irreversible, and the
first-order action variation is non-negative \citep{LiPhD2016}.}, the
strong-form governing equations for the cracking solid read
\begin{subequations}\label{eq:pde}
  \begin{equation}
    \begin{dcases}
      \nabla\cdot\bfsigma + \bforce = \rho\ddot{\disp}
      & \text{in } \varOmega, \\[4pt]
      \omega'(d)\,\overline{\mathcal{Y}}
      + \dfrac{G_{\mathrm{f}}}{c_\alpha}
      \!\left(\frac{\alpha'(d)}{\ell_o} - 2\ell_o\,\Delta d\right) \ge 0,
      \quad \dot{d} \ge 0
      & \text{in } \varOmega,
    \end{dcases}
  \end{equation}
  \text{with natural boundary conditions}
  \begin{equation}
    \begin{dcases}
      \bfsigma\cdot\Vector{n} = \sforce
      & \text{on } \partial\varOmega_{t}, \\[4pt]
      \nabla d \cdot \Vector{n} = 0
      & \text{on } \partial\varOmega,
    \end{dcases}
  \end{equation}
\end{subequations}
where $\bfsigma := \partial\psi_0^{-}/\partial\bfepsilon +
\omega(d)\,\partial\psi_0^{+}/\partial\bfepsilon$ is the Cauchy stress
tensor, $\ddot{\disp}$ is the acceleration field, $\omega'(d)$ and
$\alpha'(d)$ denote derivatives with respect to $d$,
$\overline{\mathcal{Y}} = \psi_0^{+}$ is the effective crack driving
energy, $\Delta d := \sum_i \partial^2 d / \partial x_i^2$ is the
Laplacian of the damage field, and $\Vector{n}$ is the outward unit
normal to the boundary. The inequality form of the damage evolution
equation, together with the rate constraint $\dot{d} \ge 0$ and the
bounds $0 \le d \le 1$, expresses the Karush--Kuhn--Tucker conditions
enforcing crack irreversibility. The problem is completed by the
Dirichlet conditions $\disp = \boldsymbol{u}^{*}(\boldsymbol{x})$ on
$\partial\varOmega_u$ and the initial conditions
$\boldsymbol{u}(\boldsymbol{x},0) = \boldsymbol{u}_0(\boldsymbol{x})$ and
$\dot{\boldsymbol{u}}(\boldsymbol{x},0) = \boldsymbol{v}_0(\boldsymbol{x})$.

The choice of the pair $\{\alpha(d), \omega(d)\}$ distinguishes the
principal PFM variants. The \texttt{AT2} model \cite{Bourdin2000,
miehe2010thermodynamically}, the most widely used in engineering
practice, employs $\alpha(d) = d^2$ and $\omega(d) = (1-d)^2$; however,
it lacks an elastic domain, meaning damage initiates immediately upon
any loading no matter how small. The \texttt{PF-CZM}
\cite{wuUnifiedPhasefieldTheory2017, wuLengthScaleInsensitive2018a}
generalizes this through a rational degradation function,
\begin{equation}\label{eq:pfczm-deg}
    \omega(d) = \frac{(1-d)^2}{(1-d)^2 + a_1\, d \cdot P(d)},
    \qquad
    a_1 = \frac{4}{\pi}\cdot\frac{l_\text{ch}}{\ell_o},
\end{equation}
with $\alpha(d) = 2d - d^2$ and $P(d) = 1 + a_2 d + a_2 a_3 d^2$, where
the coefficients $\{a_2, a_3\}$ are calibrated to reproduce a target
traction-separation law (TSL); see \Cref{eq:softening-laws}. Here,
$E_0$ denotes Young's modulus, $f_{\mathrm{t}}$ is the material tensile
strength, and $l_\text{ch} = E_0 G_{\mathrm{f}}/f_{\mathrm{t}}^2$
is the Irwin's characteristic length, which measures the intrinsic size of the
fracture process zone. The rational form of $\omega(d)$ introduces a
well-defined elastic threshold and enables \texttt{PF-CZM} to recover
accurate peak loads and post-peak softening responses independent of the
length scale $\ell_o$, extending its applicability to both brittle and
cohesive fracture. The two models are summarized in \Cref{tab:regime}.  The table also lists the damage profile and the damage bandwidth of each model. The damage profile is the one-dimensional solution $d(x)$ of the damage equation across a fully developed crack located at $x = 0$, obtained by minimizing the fracture energy term of \Cref{eq:lagrangian} subject to $d(0) = 1$; it describes how the damage decays with the distance $|x|$ from the crack. The damage bandwidth is the width of the zone in which $d > 0$. For \texttt{AT2} the profile decays exponentially and never reaches zero, so the bandwidth is infinite, whereas for \texttt{PF-CZM} the damage vanishes at a distance $\pi\ell_o/2$ on either side of the crack, giving a finite bandwidth $\pi\ell_o$. Neither quantity appears in the governing equations; both follow from the choice of $\alpha(d)$ and set the mesh resolution required around the crack.

\noindent In this study, two softening laws are considered for \texttt{PF-CZM}:
\begin{align}\label{eq:softening-laws}
\begin{cases}
  \text{Linear softening:}   & a_{2} = -\tfrac{1}{2},\quad a_{3} = 0, \\[4pt]
  \text{Concrete softening:} & a_{2} = 1.3868,\quad a_{3} = 0.9106.
\end{cases}
\end{align}

\subsection{Mathematical problem statement}
\label{sec:problem_statement}

The central task is to learn an operator surrogate model that maps parametrized problem inputs to the spatio-temporal evolution of the phase-field damage 
variable over the spatial domain $\Omega$. Two distinct phase-field 
formulations are employed across the two datasets, differing in both 
the mathematical structure of the phase-field variable and the 
governing energy functional.

Dataset~1 adopts the vector-valued phase-field formulation of 
\citet{hakimzadeh2022phase}, in which the damage state is described 
by a vector field $\mathbf{d}(\mathbf{x}, t) \in \mathbb{R}^2$ under 
large-deformation contact mechanics, governed by the Lagrangian of 
\Cref{eq:lagrangian} with the \texttt{AT2} choices $\alpha(d) = d^2$
and $\omega(d) = (1-d)^2$ (see \Cref{tab:regime}), where $d \equiv \|\mathbf{d}\|$
is the scalar damage measure (Euclidean norm of the vector phase field).
The surrogate predicts the full two-component vector field $\mathbf{d}(\mathbf{x}, t) \in \mathbb{R}^2$
across multiple load steps, given the initial displacement field 
$\mathbf{u}(\mathbf{x}, 0)$:
\begin{equation}
\label{eq:operator_dataset1}
    \mathcal{G}_1: \mathbf{u}(\mathbf{x}, 0) \;\mapsto\; 
    \mathbf{d}(\mathbf{x}, t), 
    \quad \mathbf{d} \in \mathbb{R}^2, 
    \quad t \in \{t_1, \ldots, t_{N_t}\}.
\end{equation}
The scalar damage indicator $\|\mathbf{d}(\mathbf{x},t)\| \in [0,1]$ is used
for evaluation and visualization throughout.

Dataset~2 adopts the scalar \texttt{PF-CZM} formulation 
\cite{wuUnifiedPhasefieldTheory2017, wuLengthScaleInsensitive2018a}, 
in which the damage state is described by a scalar phase-field 
$d(\mathbf{x}, t) \in [0, 1]$, where $d = 0$ denotes intact 
material and $d = 1$ denotes fully fractured material, governed 
by the rational degradation function of \Cref{eq:pfczm-deg} and 
solved with adaptive mesh refinement \citep{krishnan2024adaptive}. 
The surrogate predicts the scalar phase-field at the final load step, given the initial phase-field configuration, $d(\mathbf{x}, t=0)$, and the spatially 
varying Young's modulus field $E(\mathbf{x})$:
\begin{equation}
\label{eq:operator_dataset2}
    \mathcal{G}_2: \bigl(d(\mathbf{x}, 0),\, E(\mathbf{x})\bigr) 
    \;\mapsto\; d(\mathbf{x}, t_f),
    \quad d \in [0, 1],
\end{equation}
where $t_f$ denotes the final load step. In both datasets, the surrogate replaces a coupled nonlinear finite 
element solve at each load increment with a single forward pass 
through the trained network. The relevant mechanical and material properties associated with each dataset are summarized in 
\Cref{tab:regime}.

\subsubsection{Dataset 1: Square plate with a central hole}
\label{sec:dataset1}

Dataset~1 contains 1,000 high-fidelity quasi-static phase-field 
fracture simulations on a square domain $\Omega$ with a central 
circular hole \citep{T1XFF19O_2025}. The simulations were performed with a FEniCS implementation of the vector phase-field model of \citet{hakimzadeh2022phase}, using linear finite elements and a staggered (alternate minimization) solution scheme. The plate is a square of side length $L = 2.2$ (normalized units) with a central 
circular hole of radius $r_{\text{hole}} = 0.5$. Throughout, 
$\mathbf{x} = (x_1, x_2)$ denotes coordinates measured from the 
center of the plate along fixed horizontal ($x_1$) and vertical 
($x_2$) axes, so that the plate occupies $[-L/2,\,L/2]^2$ and its 
outer boundary $\partial\Omega_u$ consists of the four edges 
$x_1 = \pm L/2$ and $x_2 = \pm L/2$. The hole boundary is 
traction-free. Eight radial pre-notches of length $0.05$, equally 
spaced at $45^\circ$ intervals around the hole rim, are seeded 
through the initial phase-field and act as crack nucleation sites.

\begin{figure}[H]
    \centering
        \includegraphics[width=0.75\textwidth]{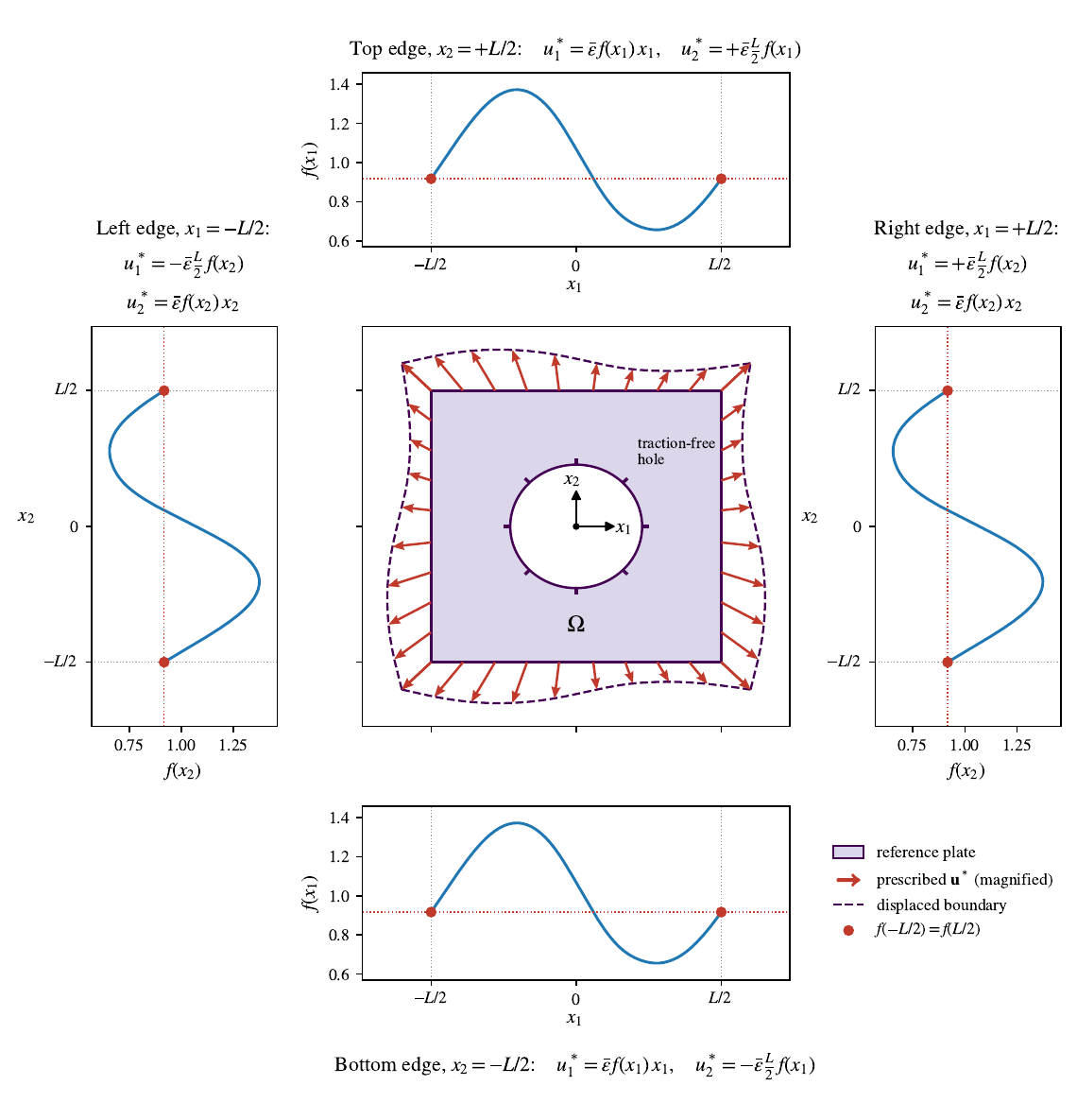}
    \caption{Boundary conditions for Dataset~1. The plate 
    occupies $[-L/2,\,L/2]^2$ with $L = 2.2$, and all panels use 
    the same plate-centered coordinates $(x_1, x_2)$. A single 
    $L$-periodic GRF realization $f(\cdot)$ is evaluated at the 
    coordinate running along each edge, $f(x_1)$ on the bottom and 
    top edges and $f(x_2)$ on the left and right edges, and the 
    prescribed displacement 
    $\mathbf{u}^{*} = \bar{\varepsilon}\,\alpha(\mathbf{x})\,\mathbf{x}$ 
    follows \Cref{eq:bc_dataset1}; the expressions next to each 
    edge are those of \Cref{eq:bc_dataset1_edges}. Arrows show the prescribed boundary displacement and the dashed curve the 
    displaced boundary, both magnified for visibility. Because 
    $f(-L/2) = f(L/2)$ (red markers), the displaced boundary is 
    continuous at the corners. The central hole, with its eight 
    radial pre-notches, is traction-free.}
    \label{BC_dataset1_details}
\end{figure}

Displacement is prescribed on the entire outer boundary as a 
spatially modulated biaxial stretch. At load step $n$,
\begin{equation}
\label{eq:bc_dataset1}
    \mathbf{u}^{*}(\mathbf{x}, t_n) 
    = \bar{\varepsilon}_n\,\alpha(\mathbf{x})\,\mathbf{x},
    \qquad
    \alpha(\mathbf{x}) = 
    \begin{cases} 
        f(x_1) & \text{on the bottom and top edges, } x_2 = \mp L/2, \\
        f(x_2) & \text{on the left and right edges, } x_1 = \mp L/2,
    \end{cases}
\end{equation}
where $\bar{\varepsilon}_n = n\,\Delta\bar{\varepsilon}$, with 
$\Delta\bar{\varepsilon} = 10^{-4}$ and $n = 1,\ldots,44$, is the 
nominal applied strain, and $f(\cdot)$ is a realization of a 
one-dimensional periodic Gaussian random field (GRF) evaluated at 
the coordinate that runs along the edge. One realization is drawn 
per simulation and used on all four edges. Substituting the edge 
coordinates into \Cref{eq:bc_dataset1} gives the explicit conditions
\begin{equation}
\label{eq:bc_dataset1_edges}
\begin{aligned}
    x_2 = \mp L/2: \qquad 
        & u_1^{*} = \bar{\varepsilon}_n\, f(x_1)\, x_1, 
        & u_2^{*} &= \mp\,\bar{\varepsilon}_n\, \tfrac{L}{2}\, f(x_1), \\
    x_1 = \mp L/2: \qquad 
        & u_1^{*} = \mp\,\bar{\varepsilon}_n\, \tfrac{L}{2}\, f(x_2), 
        & u_2^{*} &= \bar{\varepsilon}_n\, f(x_2)\, x_2,
\end{aligned}
\end{equation}
which are illustrated in \Cref{BC_dataset1_details}.

\begin{figure}[t]
    \centering
    \includegraphics[width=1\textwidth]{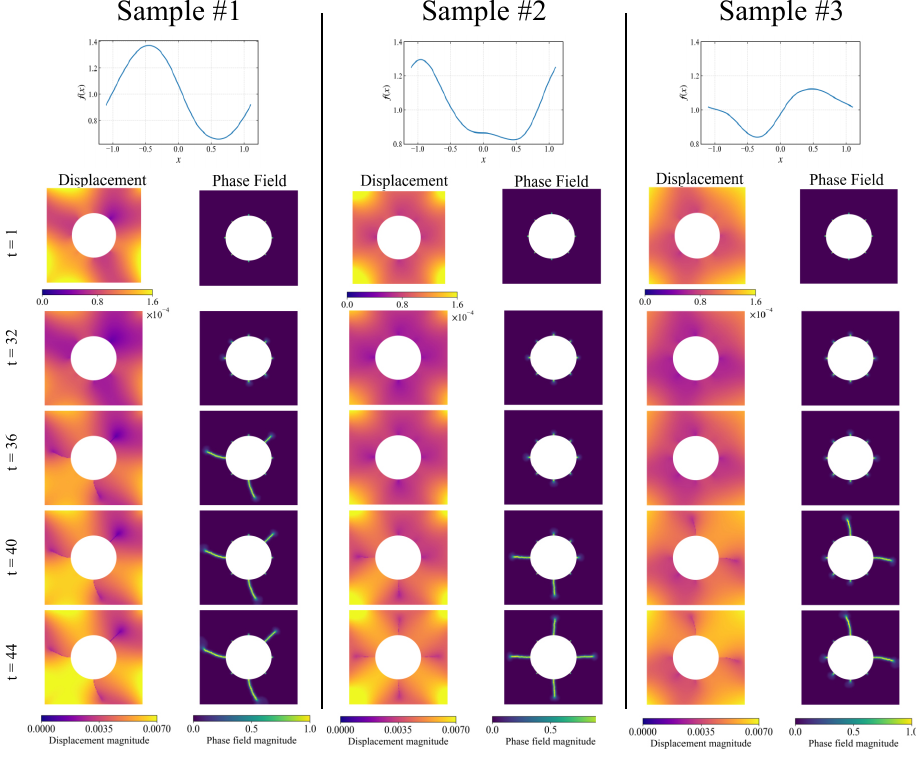}
    \caption{GRF boundary profiles and spatiotemporal 
    evolution of displacement and phase-field for three 
    representative samples. \textit{Top row:} GRF 
    realizations $f(x)$ prescribed along the boundary 
    coordinates for each sample, illustrating the diversity 
    of loading scenarios across the dataset. 
    \textit{Remaining rows:} Displacement magnitude 
    $\|\mathbf{u}(\mathbf{x}, t)\|$ (left column of each 
    sample) and phase-field magnitude 
    $\|\mathbf{d}(\mathbf{x}, t)\|$ (right column) at 
    selected load steps $t \in \{1, 32, 36, 40, 44\}$, 
    showing crack nucleation and propagation from the 
    hole boundary under GRF-driven boundary loading. }
    \label{data_snapshot}
\end{figure}

\noindent The GRF has unit mean and period $L$,
\begin{equation}
\label{eq:grf_dataset1}
    f(s) = 1 + \sum_{k=1}^{K} \left[\xi_k \cos\!\left(\frac{2\pi k s}{L}\right) 
    + \eta_k \sin\!\left(\frac{2\pi k s}{L}\right)\right],\;\;
    \xi_k,\,\eta_k \sim \mathcal{N}\!\left(0,\, s_k^2\right),\;\;
    s_k = 625\sqrt{2}\,\left[(2\pi k)^2 + 25\right]^{-2},
\end{equation}
with $K = 100$ modes and mutually independent coefficients, and 
the 1,000 simulations correspond to 1,000 independent realizations 
of $f$. Because $f$ is $L$-periodic, $f(-L/2) = f(L/2)$, so the 
prescribed displacement is single-valued at the four corners. At 
the bottom-right corner $(L/2, -L/2)$, for instance, the 
bottom-edge and right-edge expressions of 
\Cref{eq:bc_dataset1_edges} give 
$\bar{\varepsilon}_n f(L/2)\,(L/2, -L/2)$ and 
$\bar{\varepsilon}_n f(-L/2)\,(L/2, -L/2)$, respectively, which 
coincide. With unit mean and a standard deviation of approximately 
$0.21$, $f$ produces a predominantly tensile biaxial loading whose 
amplitude varies along the boundary. Each simulation stores the displacement field $\mathbf{u}(\mathbf{x}, t_n)$ and the vector phase-field $\mathbf{d}(\mathbf{x}, t_n)$ at all 44 load steps, interpolated from the finite element mesh onto a uniform $128 \times 128$ grid covering $[-L/2,\,L/2]^2$, with grid points inside the hole set to zero.
Representative snapshots of the 
GRF boundary profiles, displacement magnitude field 
$\|\mathbf{u}(\mathbf{x}, t)\|$, and the corresponding 
phase-field $\mathbf{d}(\mathbf{x}, t)$ at selected load 
steps are shown in \Cref{data_snapshot}, demonstrating 
how the damage field evolves from the initial notched state 
to the final fracture across the loading sequence.

To assess surrogate performance under a varying number of timesteps, three cases 
are evaluated: Case~1 uses 11 time steps (every 4th step from the simulated data), 
Case~2 uses 22 time steps (every 2nd step from the simulated data), and Case~3 uses all 44 time steps. 

The dataset does not prescribe a training objective, and users are free to choose their own. For the reference results in this paper, the test metric, and the training loss of every architecture except Latent DeepONet and the cDDPM (\Cref{tab:training_losses}), is a masked mean squared error (MSE)
that excludes the hole $\Omega_{\text{hole}}$, 
since grid points inside the hole carry no physical 
information. Specifically, a binary spatial mask $\mathcal{M}(\mathbf{x}) \in \{0,1\}$ is defined as $\mathcal{M}(\mathbf{x}) = \mathbf{1}[x_1^2 + x_2^2 \geq r_{\text{hole}}^2]$, where $r_{\text{hole}} = 0.5$ is the hole radius; both the training fields and the squared residuals are multiplied element-wise by $\mathcal{M}(\mathbf{x})$, so that hole-interior grid points contribute zero:
\begin{equation}
\label{eq:loss_dataset1}
    \mathcal{L}_{\text{MSE}} = \frac{1}{N_\Omega}
\sum_{\mathbf{x} \in \Omega} \mathcal{M}(\mathbf{x}) 
    \left\|\mathbf{d}_{\text{pred}}(\mathbf{x}, t) 
    - \mathbf{d}_{\text{true}}(\mathbf{x}, t)\right\|^2,
\end{equation}
where $N_\Omega = 128 \times 128$ is the total number of grid points, and $\|\cdot\|$ denotes the Euclidean norm over 
the two phase-field components.

\subsubsection{Dataset 2: Heterogeneous plate with two 
circular inclusions}
\label{sec:dataset2}
Dataset~2 contains 4,000 high-fidelity \texttt{PF-CZM} 
simulations of crack propagation in a heterogeneous
plate containing two circular inclusions 
\citep{T1RZFI3I_2025}. The growth of the defects in the domain was simulated following the formulation and 
adaptive mesh refinement strategy of 
\citet{krishnan2024adaptive}. The domain is a 
$4\,\text{mm} \times 8\,\text{mm}$ rectangular plate, 
$\Omega = [0, W] \times [0, H]$ with $W = 4$~mm and $H = 8$~mm 
and coordinates $(x_1, x_2)$ measured from the bottom-left 
corner. The two inclusions have fixed diameters of 2~mm (upper, 
centered at $(1.5, 6)$~mm) and 3~mm (lower, centered at 
$(2, 2)$~mm). The material properties for the base matrix and 
both inclusion types are listed in \Cref{tab:regime}.

The plate is loaded in Mode~I under displacement control. The 
bottom edge is fully clamped, $u_1 = u_2 = 0$ on $x_2 = 0$. On 
the top edge only the vertical displacement is prescribed, 
$u_2 = \bar{u}_n$ on $x_2 = H$, where $\bar{u}_n = n\,\Delta\bar{u}$ 
with $\Delta\bar{u} = 10^{-4}$~mm and $n = 1,\ldots,31$; the 
horizontal displacement on the top edge is left free (zero shear 
traction). The left and right edges, $x_1 = 0$ and $x_1 = W$, are 
traction-free.

\begin{figure}[H]
    \centering
        \includegraphics[width=1.0\textwidth]{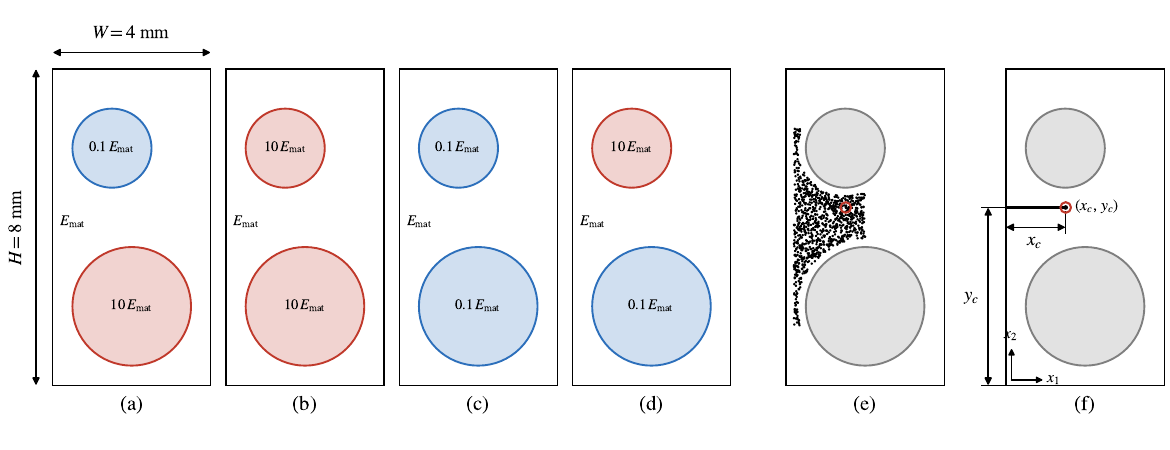}
        \caption{Inclusion configurations and initial cracks in 
    Dataset~2. The plate measures $4\,\text{mm} \times 8\,\text{mm}$. 
    Blue circles denote less stiff inclusions 
    ($E_{\text{inc}} = 0.1\,E_{\text{mat}}$), which tend to 
    attract and channel the propagating crack; red circles 
    denote stiffer inclusions 
    ($E_{\text{inc}} = 10\,E_{\text{mat}}$), which act to 
    deflect the crack around their boundary (see 
    \Cref{tab:regime}). The four inclusion 
    configurations, labeled as \textit{smaller inclusion, 
    larger inclusion}, are (a) less stiff, stiffer; 
    (b) stiffer, stiffer; (c) less stiff, less stiff; and 
    (d) stiffer, less stiff. (e) The 1,000 crack-tip seeds 
    (black dots), which lie in the left half of the plate, 
    away from both inclusions and from the top and bottom 
    edges. (f) One representative seed (red circle in (e) and 
    (f)) and its horizontal pre-crack (thick black line), 
    which runs from the left edge to the crack tip 
    $(x_c, y_c)$ and constitutes the initial condition 
    $d(\mathbf{x}, 0)$ of the simulation. In (e) and (f) the 
    inclusions are drawn in gray because the seed locations 
    are generated independently of the inclusion configuration.}

    \label{fig:points}
\end{figure}

Two inputs vary across the simulations: the inclusion 
configuration and the initial crack. Each inclusion is assigned, 
independently, a material that is either stiffer or less stiff 
than the base matrix, yielding four distinct inclusion 
configurations illustrated in \Cref{fig:points}(a-d). 
Stiffer inclusions ($E_{\text{inc}} = 10\,E_{\text{mat}}$) deflect 
the propagating crack around their boundary, while less stiff 
inclusions ($E_{\text{inc}} = 0.1\,E_{\text{mat}}$) attract and 
guide the crack through them, where $E_{\text{mat}}$ denotes the 
Young's modulus of the base matrix. The material 
properties for the base matrix and both inclusion types 
are listed in \Cref{tab:regime}.

A set of 1,000 candidate crack-tip locations is generated 
randomly within the domain, constrained such that the 
$x$-coordinate lies within the left half of the domain 
and the crack-tip is sufficiently far from both inclusions 
and the top and bottom boundaries, ensuring each crack can 
propagate freely toward the inclusions under the applied 
loading. The generation procedure is illustrated in 
\Cref{fig:points}: panel~(e) shows all 1,000 admissible 
crack-tip seeds distributed across the left half of the 
domain, and panel~(f) shows a single representative 
seed together with its initialized horizontal pre-crack 
$d(\mathbf{x}, 0)$, which serves as the initial 
condition for the corresponding simulation as well as the surrogate models. For each of the four inclusion configurations, the same 1,000 initial 
cracks are used, yielding a total of 4,000 simulations. 

\begin{figure}[H]
    \centering
    \includegraphics[width=1.0\textwidth]{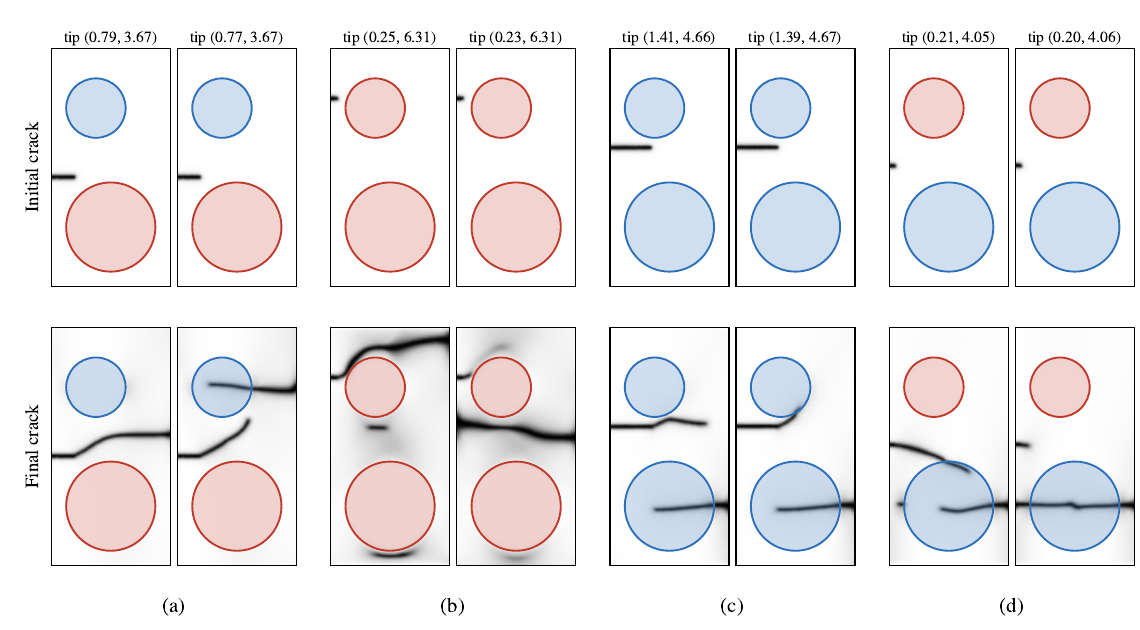}
    \caption{Sensitivity of the crack path to the initial crack 
    location in Dataset~2. Each panel shows two simulations with 
    the \emph{same} inclusion configuration, labeled as 
    \textit{smaller inclusion, larger inclusion}: (a) less stiff, 
    stiffer; (b) stiffer, stiffer; (c) less stiff, less stiff; 
    and (d) stiffer, less stiff. Within each pair the initial 
    crack tips (coordinates in mm above each column) are about 
    $0.02$~mm apart. \textit{Top row:} initial phase-field 
    $d(\mathbf{x}, 0)$. \textit{Bottom row:} phase-field 
    $d(\mathbf{x}, t_f)$ at the final load step $t_f = 31$. Blue 
    and red circles mark less stiff and stiffer inclusions, 
    respectively. Despite nearly identical inputs, the two 
    simulations of each pair end in qualitatively different crack 
    patterns.}
   
    \label{fig:challenging_dataset2}
\end{figure}

The crack path can be highly sensitive to the initial crack 
location, even when the inclusions are unchanged. 
\Cref{fig:challenging_dataset2} shows four pairs of simulations, 
one pair per inclusion configuration. Within each pair the 
inclusions are identical and the initial crack tips are about 
$0.02$~mm apart, which is less than one cell of the 
$128 \times 256$ grid, yet the final crack patterns differ 
qualitatively: the pre-existing crack crosses the plate between 
the inclusions, is drawn toward an inclusion, or arrests while a 
new crack nucleates elsewhere. Such divergence is not the rule. 
For most pairs of neighboring initial cracks the final paths 
nearly coincide, and only a few percent of the pairs whose tips 
lie within $0.1$~mm of each other end in clearly different crack 
patterns. These near-bifurcation cases are nevertheless what makes 
the prediction task difficult, because the surrogate must resolve 
differences in the input that are smaller than one grid cell.

Each simulation stores the displacement field and the scalar 
phase-field at all 31 load steps, together with the Young's 
modulus field $E(\mathbf{x})$, interpolated from the adaptively 
refined finite element mesh onto a uniform $128 \times 256$ grid.

The benchmark task maps the initial phase-field and the modulus 
field to the phase-field at the final load step, 
\Cref{eq:operator_dataset2}; the intermediate load steps and the 
displacement fields remain available for other tasks 
(\Cref{sec:extending}). The dataset is split into 3,600 training 
and 400 test samples (see \Cref{tab:regime}), that is, 
900 and 100 per inclusion configuration. As for Dataset~1, the 
training objective is the user's choice, and the losses used for the 
reference results are listed in \Cref{tab:training_losses}. All 
architectures are scored on the test set with the MSE over the full domain:
\begin{equation}
\label{eq:loss_dataset2}
    \mathcal{L}_{\text{MSE}} = \frac{1}{N_\Omega} 
    \sum_{\mathbf{x} \in \Omega} 
    \left(d_{\text{pred}}(\mathbf{x}, t_f) 
    - d_{\text{true}}(\mathbf{x}, t_f)\right)^2,
\end{equation}
where $N_\Omega = 128 \times 256$ is the total number of grid points and 
$t_f = 31$ denotes the final load step.

\subsection{Surrogate model architectures}
\label{sec:architectures}

Five surrogate architectures are benchmarked, spanning two 
broad categories: (i) \textit{neural operators}, which learn 
mappings between infinite-dimensional function spaces: Fourier neural operator (FNO), deep operator network (DeepONet), and Latent DeepONet, and (ii) \textit{deep 
generative and convolutional models}, which treat the 
prediction as a structured image-to-image regression or 
conditional generation task: UNet and a Conditional 
Denoising Diffusion Probabilistic Model (cDDPM). Each 
architecture receives the same input fields, is trained on the same 
data split within the same computational budget, and is evaluated on 
the same test set with the same metric, so that the reported errors 
are directly comparable. The training losses are not identical across 
architectures; they are listed in \Cref{tab:training_losses} and 
explained in \Cref{sec:training}. These choices define the reference protocol of this study and are not a constraint of the benchmark: users may adopt other losses, preprocessing, or training budgets, provided the results are reported with the common test metrics of \Cref{sec:training}.

\subsubsection{Fourier neural operator (FNO)}
\label{sec:fno}

The Fourier neural operator \citep{li2020fourier} learns a 
mapping between function spaces by parameterizing the integral 
kernel in Fourier space, enabling discretization-invariant 
predictions on functions defined over continuous domains. Each 
FNO layer applies a global convolution in Fourier space combined 
with a local linear transformation:
\begin{equation}
\label{eq:fno_layer}
    \mathbf{v}^{(\ell_o+1)}(\mathbf{x}) = \sigma\!\left(
    W^{(\ell_o)} \mathbf{v}^{(\ell_o)}(\mathbf{x}) + 
    \mathcal{F}^{-1}\!\left(R_\theta^{(\ell_o)} \cdot 
    \mathcal{F}\!\left(\mathbf{v}^{(\ell_o)}\right)
    \right)\!(\mathbf{x})\right),
\end{equation}
where $\mathbf{v}^{(\ell_o)}$ is the feature map at layer $\ell_o$, 
$W^{(\ell_o)}$ is a pointwise linear transformation, 
$\mathcal{F}$ and $\mathcal{F}^{-1}$ denote the forward and 
inverse Fourier transforms, respectively, $R_\theta^{(\ell_o)}$ is a learnable weight tensor truncated to the lowest Fourier modes, and 
$\sigma$ is the activation function. The input is first lifted 
to a high-dimensional feature space by a shallow MLP, passed 
through $L$ successive FNO layers, and then projected to the 
output space by a second shallow MLP.

\begin{figure}[H]
    \centering
    \includegraphics[width=\textwidth]{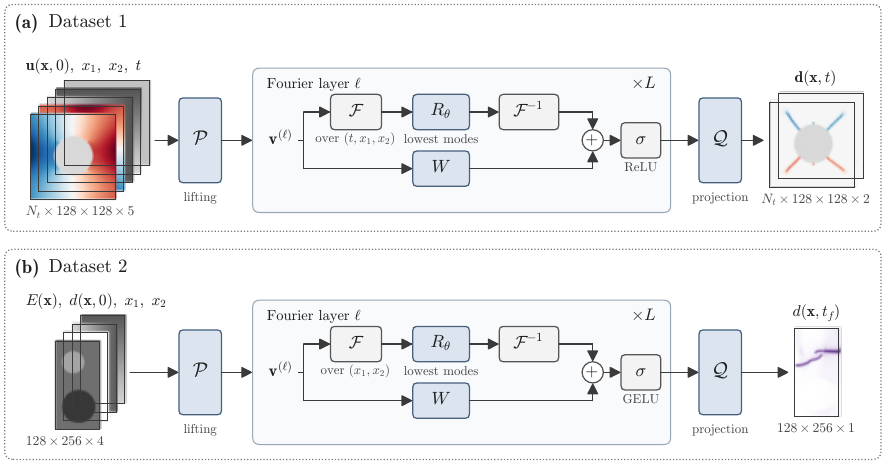}
    \caption{\textbf{FNO architecture} (a) Dataset~1 : The initial 
    displacement field $\mathbf{u}(\mathbf{x}, 0)$ is 
    broadcast across $N_t$ time steps and concatenated with 
    spatial coordinates $(x, y)$ and normalized time $t$ 
    to form a five-channel spatiotemporal input. After lifting 
    to a high-dimensional feature space, $L$ Fourier layers 
    apply global convolutions jointly across the spatial and 
    temporal dimensions via \Cref{eq:fno_layer}. The output 
    is the two-channel vector phase-field 
    $\mathbf{d}(\mathbf{x}, t) \in \mathbb{R}^2$ over all 
    $N_t$ load steps; 
    (b) Dataset~2 : The initial 
    scalar phase-field $d(\mathbf{x}, 0)$ and Young's modulus 
    field $E(\mathbf{x})$ are concatenated with spatial 
    coordinates $(x, y)$ to form a four-channel spatial 
    input. After lifting, $L$ Fourier layers apply global 
    convolutions across the two spatial dimensions only. 
    The output is the single-channel scalar phase-field 
    $d(\mathbf{x}, t_f)$ at the final load step.}
    \label{fig:fno_d1}
\end{figure}

For \textbf{Dataset~1}, the problem requires predicting the 
spatiotemporal evolution of the vector phase-field 
$\mathbf{d}(\mathbf{x}, t)$ over $N_t$ load steps given the 
initial displacement field $\mathbf{u}(\mathbf{x}, 0)$. 
The input tensor is constructed by broadcasting 
$\mathbf{u}(\mathbf{x}, 0)$ across all time steps and 
appending the spatial coordinates $(x_1, x_2)$ and the 
normalized time coordinate $t$ as additional channels, 
yielding a five-channel spatiotemporal input of shape 
$(N_t \times 128 \times 128 \times 5)$. A three-dimensional 
\texttt{FNO3D} is used, with Fourier convolutions applied 
jointly across the two spatial dimensions and the temporal 
dimension, and ReLU activations. The output is the 
two-channel vector phase-field $\mathbf{d}(\mathbf{x}, t)$ 
of shape $(N_t \times 128 \times 128 \times 2)$. The masked 
MSE loss of \Cref{eq:loss_dataset1} is used during training, 
excluding the hole region $\Omega_{\text{hole}}$. 
The Dataset~1 FNO architecture is illustrated in 
\Cref{fig:fno_d1}(a).

For \textbf{Dataset~2}, the problem reduces to a static 
state-to-state mapping: given the initial scalar phase-field 
$d(\mathbf{x}, 0)$ and the spatially-varying Young's modulus  $E(\mathbf{x})$, 
predict the scalar phase-field $d(\mathbf{x}, t_f)$ at the 
final load step, $t_f$. The two input fields are concatenated as a 
four-channel spatial tensor of shape 
$(128 \times 256 \times 4)$, where the additional two channels 
are the spatial coordinates $(x_1, x_2)$. Since the temporal 
dimension is absent, we employ \texttt{FNO2D}, with Fourier convolutions restricted to the spatial 
dimensions and GELU activations. The output is the single-channel 
scalar phase-field $d(\mathbf{x}, t_f)$ of shape 
$(128 \times 256 \times 1)$. The weighted MAE of \Cref{eq:wmae_dataset2} is used as the training loss.  
The Dataset~2 FNO architecture is illustrated in \Cref{fig:fno_d1}(b).

\subsubsection{Deep operator network (DeepONet)}
\label{sec:deeponet}
DeepONet \citep{lu2019deeponet} is grounded in the universal 
approximation theorem for operators \citep{chen1995universal} and decomposes the target operator into two subnetworks: a 
branch network that encodes the input function evaluated at a 
fixed set of sensor points, $\mathbf{x}_s$, and a trunk network that encodes 
the continuous query coordinates at which the output is 
evaluated. The predicted output at query location 
$(\mathbf{x}, t)$ is given by:
\begin{equation}
\label{eq:deeponet}
    \mathcal{G}(\mathbf{u})(\mathbf{x}, t) 
    = \sum_{k=1}^{p} b_k\!\left(\mathbf{u}(\mathbf{x}_s, 0)
    \right) \cdot \tau_k(\mathbf{x}, t) + b_0,
\end{equation}
where $b_k$ are the $p$ embeddings of the branch network for 
the input function, $\tau_k$ are 
the $p$ embeddings of the trunk network encoding the query 
location $(\mathbf{x}, t)$, and $b_0$ is a learnable bias. 
The branch and trunk outputs are combined via a dot product, 
making the architecture highly flexible for operator learning 
across arbitrary query locations.

\begin{figure}[H]
    \centering
    \includegraphics[width=\textwidth]{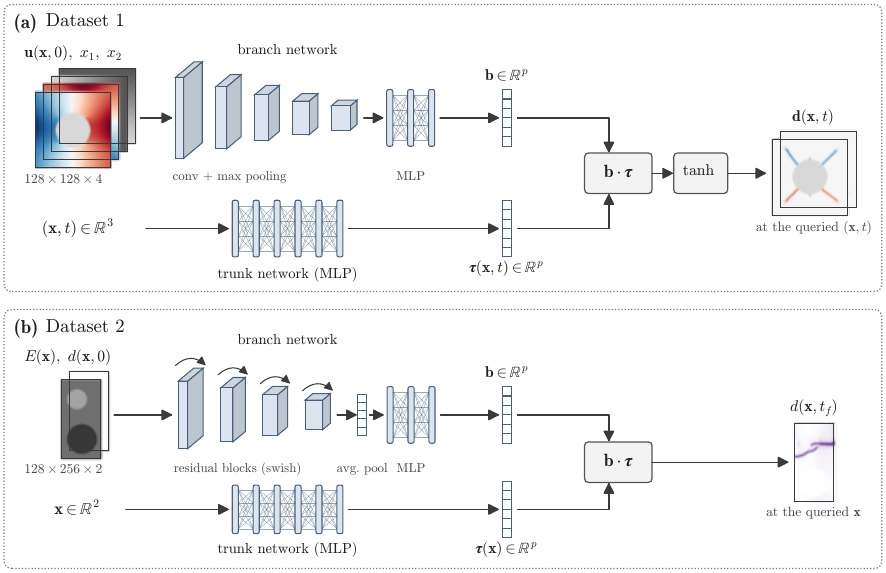}
    \caption{\textbf{DeepONet architecture} (a) Dataset~1 : The branch 
    network encodes the initial displacement field 
    $\mathbf{u}(\mathbf{x}, 0)$ concatenated with spatial 
    coordinates via a CNN--MLP, producing $p$ latent 
    coefficients $\{b_k\}$. The trunk network encodes the 
    spatiotemporal query $(\mathbf{x}, t) \in \mathbb{R}^3$ 
    via an MLP, producing $p$ basis functions $\{\tau_k\}$. 
    Their dot product gives the two-channel vector phase 
    field $\mathbf{d}(\mathbf{x}, t) \in \mathbb{R}^2$ 
    at the queried locations; (b) Dataset~2: The branch 
    network encodes the concatenated initial scalar phase 
    field $d(\mathbf{x}, 0)$ and Young's modulus field 
    $E(\mathbf{x})$ via a ResNet-style CNN, producing $p$ 
    latent coefficients $\{b_k\}$. The trunk network encodes 
    purely spatial query coordinates $\mathbf{x} \in 
    \mathbb{R}^2$ via an MLP, producing $p$ basis functions 
    $\{\tau_k\}$. Their dot product gives the single-channel 
    scalar phase-field $d(\mathbf{x}, t_f)$ at the final 
    load step.}
    \label{fig:don_d1}
\end{figure}

For \textbf{Dataset~1}, the branch network processes the 
initial displacement field $\mathbf{u}(\mathbf{x}, 0)$ 
concatenated with the spatial coordinates $(x_1, x_2)$, 
yielding a four-channel input of shape 
$(128 \times 128 \times 4)$. The branch network is a CNN 
with max pooling followed by an MLP, encoding the spatial 
structure of the input into a latent vector of dimension $p$. 
The trunk network is an MLP that takes the spatiotemporal 
query coordinates $(\mathbf{x}, t) \in \mathbb{R}^3$ as 
input. The branch and trunk outputs are combined via the dot 
product of \Cref{eq:deeponet} followed by a $\tanh$ 
activation, and the output is the two-channel vector phase 
field $\mathbf{d}(\mathbf{x}, t) \in \mathbb{R}^2$ evaluated 
at the queried spatiotemporal locations. The masked MSE loss 
of \Cref{eq:loss_dataset1} is applied during training. The DeepONet architecture corresponding to Dataset~1 is illustrated in 
\Cref{fig:don_d1}(a).

For \textbf{Dataset~2}, the temporal dimension is removed 
entirely. The branch network encodes the concatenated initial 
scalar phase-field $d(\mathbf{x}, 0)$ and Young's modulus 
field $E(\mathbf{x})$ as a two-channel input of shape 
$(128 \times 256 \times 2)$, using a ResNet-style CNN, swish activations, and average pooling 
to better capture the multi-scale spatial structure of the 
inclusion configurations. The trunk network takes only 
spatial coordinates $(\mathbf{x}) \in \mathbb{R}^2$ as 
input. Since our desired phase-field corresponds to a single time step (final), the temporal coordinates are not queried. The 
output is passed directly without a final nonlinear 
activation, and the loss is a weighted mean absolute error 
(MAE) that penalizes errors in the damaged region 
($d > 0.5$) more heavily to account for the class imbalance 
between intact and fractured material. The output is the 
single-channel scalar phase-field $d(\mathbf{x}, t_f)$ at 
the final load step. The DeepONet architecture for Dataset~2 
is illustrated in \Cref{fig:don_d1}(b).

\subsubsection{Latent DeepONet}
\label{sec:latent_deeponet}

Latent DeepONet \citep{kontolati2023learning} extends the 
standard DeepONet framework by operating in a compressed 
latent space, reducing the dimensionality of the prediction 
target before learning the neural operator. A convolutional autoencoder is first trained to learn a low-dimensional latent 
representation of the high-dimensional output field. DeepONet 
is then trained to predict directly in this latent space, 
with the decoder subsequently mapping latent predictions back 
to the physical domain. The encoder 
and decoder are pretrained and the trainable parameters are frozen before DeepONet 
training begins.

\begin{figure}[H]
    \centering
    \includegraphics[width=\textwidth]{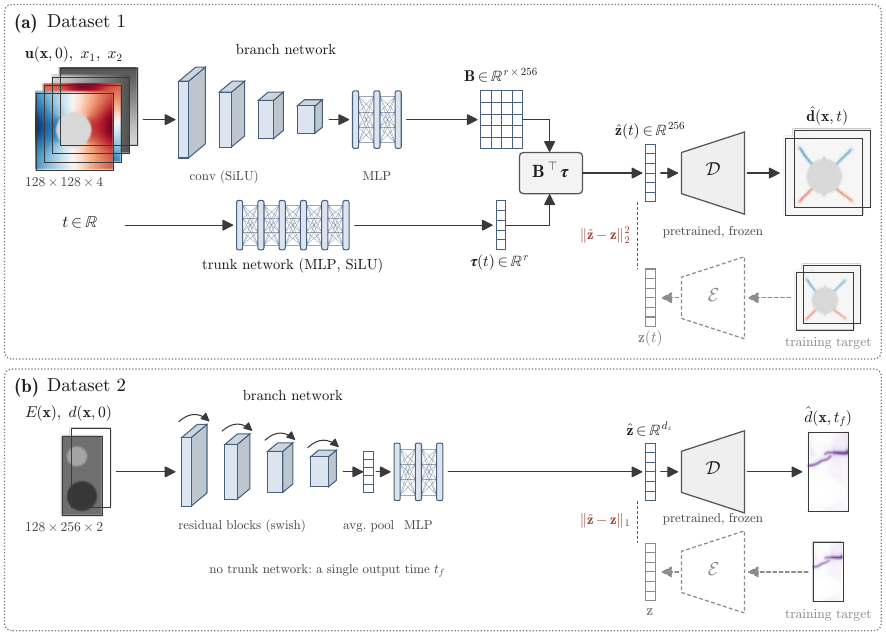}
    \caption{\textbf{Latent DeepONet architecture} (a) Dataset~1: 
    A pretrained convolutional autoencoder compresses the 
    two-channel vector phase-field $\mathbf{d}(\mathbf{x}, t)$ 
    into a latent sequence $\mathbf{z}(t) \in \mathbb{R}^{256}$. 
    The branch network encodes $\mathbf{u}(\mathbf{x}, 0)$ 
    and spatial coordinates via a CNN--MLP; the trunk network 
    encodes only the normalized time $t$. Their combined output 
    predicts $\hat{\mathbf{z}}(t)$, which is decoded to 
    recover $\hat{\mathbf{d}}(\mathbf{x}, t)$; 
    (b) Dataset~2: A pretrained convolutional autoencoder 
    compresses the scalar phase-field $d(\mathbf{x}, t_f)$ into 
    a latent vector $\mathbf{z} \in \mathbb{R}^{d_z}$. The trunk 
    network is bypassed entirely since there is no temporal query 
    coordinate. The branch network encodes the concatenated inputs 
    $d(\mathbf{x}, 0)$ and $E(\mathbf{x})$ via a ResNet-style CNN, 
    producing a predicted latent vector $\hat{\mathbf{z}}$ that is 
    decoded to recover the scalar phase-field 
    $\hat{d}(\mathbf{x}, t_f)$ at the final load step.}
    \label{fig:ldon_d1}
\end{figure}

\noindent For \textbf{Dataset~1}, let the two-channel vector phase-field solution be denoted by $\mathbf{d}(\mathbf{x},t):\Omega\times[0,T]\rightarrow\mathbb{R}^{2}$, where $\mathbf{x}=(x_1,x_2)\in\Omega\subset\mathbb{R}^2$. A convolutional autoencoder is first trained to learn an encoder $\mathcal{E}$ and decoder $\mathcal{D}$, where the encoder maps the phase-field solution to a latent representation $\mathbf{z}(t)=\mathcal{E}(\mathbf{d}(\cdot,t))\in\mathbb{R}^{256}$ and the decoder reconstructs the physical solution as $\mathbf{d}(\cdot,t)\approx\mathcal{D}(\mathbf{z}(t))$. The branch network receives as input the initial displacement field $\mathbf{u}(\mathbf{x},0)$ concatenated with the spatial coordinates $(x_1,x_2)$, forming the four-channel input $\mathbf{b}(\mathbf{x})=[u_1(\mathbf{x},0),u_2(\mathbf{x},0),x_1,x_2]$. A CNN--MLP architecture with SiLU activations maps this input to a branch feature tensor $\mathbf{B}=\mathcal{B}(\mathbf{b})\in\mathbb{R}^{r\times256}$. The trunk network takes only the normalized temporal coordinate $t\in\mathbb{R}$ as input and produces the feature vector $\mathbf{T}(t)=\mathcal{T}(t)\in\mathbb{R}^{r}$, as the spatial dependence has already been encoded through the latent representation. The latent prediction is obtained through the DeepONet bilinear pairing $\hat{\mathbf{z}}(t)=\sum_{k=1}^{r}\mathbf{b}_{k,:}\tau_k(t)\in\mathbb{R}^{256}$, which is implemented using an Einstein summation (einsum) operation over the shared feature dimension. The predicted latent state is subsequently decoded using the pre-trained decoder to obtain the phase-field prediction in the physical domain, $\hat{\mathbf{d}}(\mathbf{x},t)=\mathcal{D}(\hat{\mathbf{z}}(t))$. The operator is trained by minimizing the MSE between the predicted and the encoded latent states, $\hat{\mathbf{z}}(t)$ and $\mathbf{z}(t)$. The decoder is applied only afterwards, and the test error is evaluated in physical space with the masked MSE of \Cref{eq:loss_dataset1}. The complete Dataset~1 Latent DeepONet architecture is illustrated in \Cref{fig:ldon_d1}(a).
For an output field 
$\mathbf{d}(\mathbf{x}, t) \in \mathbb{R}^{D_x \times D_y}$, 
the three-stage mapping is:
\begin{align}
\label{eq:ldon}
    \mathcal{E}_{\theta_{\text{enc}}} 
        &: \mathbf{d}(\mathbf{x}, t) 
        \;\mapsto\; \mathbf{z}(t) \in \mathbb{R}^{d_z}, 
        \nonumber \\
    \mathcal{G}_\theta 
        &: \mathbf{u}(\mathbf{x}, 0) 
        \;\mapsto\; \mathbf{z}(t) \in \mathbb{R}^{d_z}, 
        \\
    \mathcal{D}_{\theta_{\text{dec}}} 
        &: \mathbf{z}(t) 
        \;\mapsto\; \mathbf{d}(\mathbf{x}, t), 
        \nonumber
\end{align}
where $\mathcal{E}_{\theta_{\text{enc}}}$ is the encoder, 
$\mathcal{G}_\theta$ is the DeepONet operator trained in 
latent space, $\mathcal{D}_{\theta_{\text{dec}}}$ is the 
decoder, $\mathbf{z}(t)$ is the latent vector, and 
$d_z$ is the latent dimension, where $d_z\ll D_x \times D_y$.

For \textbf{Dataset~2}, the Latent DeepONet reduces to a purely spatial 
mapping with no temporal component, since the prediction target is a 
single spatial field at the final load step rather than a sequence. A 
convolutional autoencoder is trained to compress the scalar phase-field 
$d(\mathbf{x}, t_f) \in \mathbb{R}^{128 \times 256}$ into a 
low-dimensional latent vector $\mathbf{z} \in \mathbb{R}^{d_z}$, where 
$d_z \ll 128 \times 256$. The trunk network, having no continuous 
temporal query coordinates in a static mapping, is bypassed entirely. 
The branch network alone maps the concatenated inputs 
$\bigl(d(\mathbf{x}, 0),\, E(\mathbf{x})\bigr)$, encoded as a 
two-channel spatial tensor of shape $(128 \times 256 \times 2)$, 
directly to a predicted latent vector $\hat{\mathbf{z}} \in 
\mathbb{R}^{d_z}$ via a ResNet-style CNN with residual shortcuts and 
swish activations. The decoder $\mathcal{D}_{\theta_{\text{dec}}}$ 
subsequently maps $\hat{\mathbf{z}}$ back to the physical domain, 
recovering the scalar phase-field $\hat{d}(\mathbf{x}, t_f)$ at the 
final load step. The network is trained by minimizing the mean absolute error between 
the predicted and the encoded latent vectors, and the test error is 
evaluated on the decoded field with the MSE of \Cref{eq:loss_dataset2}. The Dataset~2 
Latent DeepONet architecture is illustrated in \Cref{fig:ldon_d1}(b).

\subsubsection{UNet}
\label{sec:unet}

\noindent UNet \citep{ronneberger2015u} is employed as a fully convolutional approximation of the nonlinear solution operator between discretized field variables. Let $\mathcal{U}_\ell$ and $\mathcal{D}_\ell$ denote the encoder and decoder operators at resolution level $\ell$, respectively. The network constructs a hierarchy of latent feature representations through successive restriction operations in the encoder and reconstructs the output through corresponding prolongation operations in the decoder. Skip connections introduce identity mappings between encoder and decoder feature spaces at matching resolutions, allowing the reconstructed solution to simultaneously incorporate coarse-scale contextual information and fine-scale spatial features.

\begin{figure}[H]
    \centering
    \includegraphics[width=\textwidth]{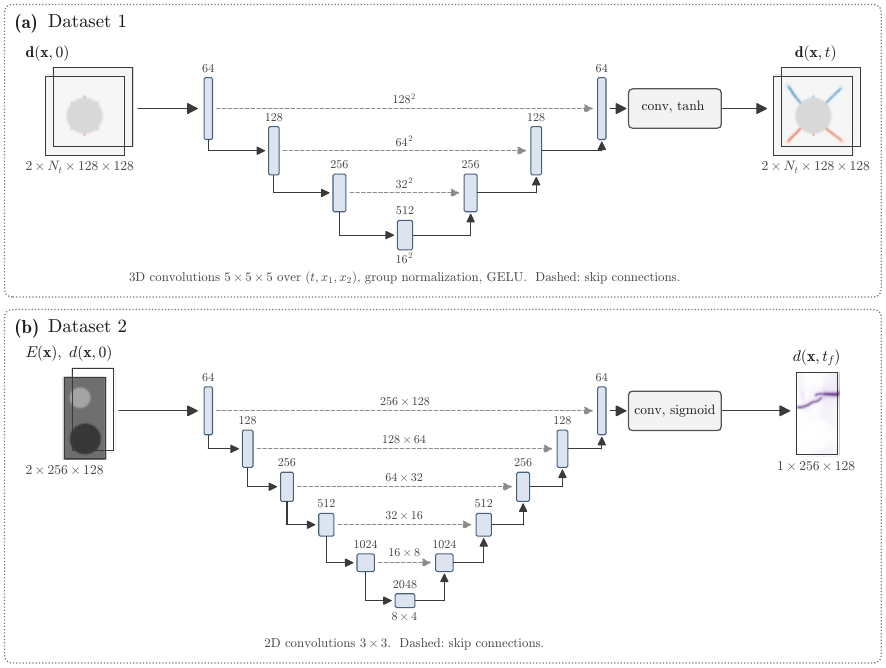}
    \caption{\textbf{UNet architecture} (a)
Dataset~1: Three 
    encoding levels with $5 \times 5 \times 5$ 
    convolutions, group normalization, and GELU 
    activations process the joint spatiotemporal input. 
    Skip connections concatenate encoder feature maps 
    into the decoder at each resolution level to 
    preserve spatial detail. Input and output shape: 
    $2 \times N_t \times 128 \times 128$, where the two 
    channels correspond to the components of the vector 
    phase-field $\mathbf{d}(\mathbf{x}, t)$ and $N_t \in 
    \{11, 22, 44\}$; (b) Dataset~2: Five 
    encoding levels with $3 \times 3$ convolutions and 
    channel widths $[64, 128, 256, 512, 1024, 2048]$ 
    process the static spatial input. A sigmoid 
    activation at the output layer constrains predictions 
    to $[0, 1]$. Input shape: $2 \times 256 \times 128$ 
    (initial phase-field $d(\mathbf{x}, 0)$ and 
    $E(\mathbf{x})$); output shape: $1 \times 256 \times 
    128$ (scalar phase-field $d(\mathbf{x}, t_f)$).}
    \label{fig:3dunet}
\end{figure}

\medskip
\noindent\textbf{Dataset~1.}
For Dataset~1, the objective is to learn the mapping
$
\mathcal{G}_1:\mathbf{d}(\mathbf{x},0)\mapsto \mathbf{d}(\mathbf{x},t),
$
for all discrete loading steps $t\in\{t_1,\ldots,t_{N_t}\}$, where $\mathbf{d}(\mathbf{x},t)\in\mathbb{R}^2$ denotes the vector phase-field. Rather than predicting each load step independently, the entire space--time solution trajectory is represented as a tensor
$
\mathbf{D}\in\mathbb{R}^{2\times N_t\times128\times128},
$
with channel, temporal, and spatial dimensions corresponding to the two phase-field components, the loading sequence, and the spatial discretization, respectively. A three-dimensional UNet \citep{cciccek20163d} is therefore used to approximate the operator
$
\mathcal{G}_1:\mathbb{R}^{2\times N_t\times128\times128}
\rightarrow
\mathbb{R}^{2\times N_t\times128\times128},
$
treating the temporal coordinate as an additional dimension of the computational domain. The architecture consists of three encoder-decoder levels with convolutional kernels of size $5\times5\times5$, feature widths $[64,128,256,512]$, group normalization with 32 groups, and GELU nonlinearities. The final layer applies a pointwise $\tanh$ activation so that the predicted phase-field remains bounded in $[-1,1]$. For the shortest loading sequence ($N_t=11$), the encoder depth is restricted to three levels because additional downsampling in the temporal direction would require excessive padding, introducing artificial boundary effects in the latent representation. Consequently, increasing $N_t$ provides a larger effective temporal domain for hierarchical feature extraction, which contributes to the observed improvement in predictive accuracy. The network parameters are optimized by minimizing the masked mean-squared error defined in \Cref{eq:loss_dataset1}. The complete Dataset~1 architecture is shown in \Cref{fig:3dunet}.

\medskip
\noindent\textbf{Dataset~2.}
For Dataset~2, the task is to learn the static operator
$
\mathcal{G}_2:\bigl(d(\mathbf{x},0),E(\mathbf{x})\bigr)
\mapsto
d(\mathbf{x},t_f),
$
where $d(\mathbf{x},0)\in[0,1]$ is the initial scalar phase-field, $E(\mathbf{x})$ is the spatially varying Young's modulus field, and $d(\mathbf{x},t_f)\in[0,1]$ denotes the final damage state. Since no temporal dimension is present, the operator acts solely on spatial fields defined over the two-dimensional domain $\Omega$.

The network input is represented by the tensor
$
\mathbf{X}\in\mathbb{R}^{2\times256\times128},
$
formed by concatenating $d(\mathbf{x},0)$ and $E(\mathbf{x})$ along the channel dimension, while the output is
$
\hat{\mathbf{Y}}\in\mathbb{R}^{1\times256\times128},
$
corresponding to the predicted final phase-field. A two-dimensional UNet is used to approximate the mapping
$
\mathcal{G}_2:\mathbb{R}^{2\times256\times128}
\rightarrow
\mathbb{R}^{1\times256\times128}.
$
The architecture employs five encoder-decoder levels with $3\times3$ convolutions and feature widths $[64,128,256,512,1024,2048]$. A pointwise sigmoid activation is applied at the output layer to enforce the physically admissible range $[0,1]$ for the phase-field variable. Training is performed using the mean-squared error loss defined in \Cref{eq:loss_dataset2}. The Dataset~2 UNet architecture is illustrated in \Cref{fig:3dunet}.

\subsubsection{Conditional denoising diffusion probabilistic model}
\label{sec:diffusion}

The cDDPM \citep{ho2020denoising} frames the surrogate 
prediction task as a conditional generative problem. A 
forward Markov chain of $T$ steps gradually corrupts the 
ground truth output field with isotropic Gaussian noise:
\begin{equation}
\label{eq:ddpm_forward}
    q(\mathbf{x}_t \mid \mathbf{x}_{t-1}) = 
    \mathcal{N}\!\left(\mathbf{x}_t;\, 
    \sqrt{1 - \beta_t}\,\mathbf{x}_{t-1},\, 
    \beta_t \mathbf{I}\right),
\end{equation}
where $\mathbf{x}_t$ is the noised field at diffusion step $t$, 
$\mathbf{x}_{t-1}$ is the field at the previous step, 
$\mathcal{N}(\cdot\,;\boldsymbol{\mu},\boldsymbol{\Sigma})$ denotes a 
Gaussian distribution with mean $\boldsymbol{\mu}$ and covariance 
$\boldsymbol{\Sigma}$, $\beta_t \in (0,1)$ is the variance (noise) 
schedule controlling the amount of noise injected at each of the 
$T$ forward steps, and $\mathbf{I}$ is the identity matrix. The model 
learns to reverse this process conditioned on the problem inputs, 
recovering the clean output field from pure noise. A cosine noise schedule is adopted in place of 
a linear one, maintaining a higher signal-to-noise 
ratio for longer during the forward process and 
enabling the network to learn semantically meaningful 
features during the early reverse steps. The training 
objective minimizes the $\ell_1$ loss between the true 
injected noise $\boldsymbol{\epsilon}$ and the 
network's noise estimate 
$\boldsymbol{\epsilon}_\theta$:
\begin{equation}
\label{eq:ddpm_loss}
    \mathcal{L}_{\text{DDPM}} = 
    \mathbb{E}_{t,\mathbf{x}_0, 
    \boldsymbol{\epsilon}}\!\left[
    \left\|\boldsymbol{\epsilon} - 
    \boldsymbol{\epsilon}_\theta\!\left(
    \mathbf{x}_t, t, \mathbf{c}\right)
    \right\|_1\right],
\end{equation}
where $\mathbf{c}$ denotes the conditioning input 
and $\mathbf{x}_0$ is the clean target field. The 
noise prediction network is a UNet backbone with 
ResNet blocks equipped with weight-standardized 
convolutions \citep{qiao2019weight}, which normalize 
convolutional kernel weights to zero mean and unit 
variance. Combined with group normalization (8 groups), 
this stabilizes training without requiring large batch 
sizes. To manage computational cost at high spatial 
resolution, a hybrid attention strategy is employed: 
linear attention with $\mathcal{O}(N)$ complexity in 
the high-resolution encoder and decoder paths, and 
full $\mathcal{O}(N^2)$ self-attention at the 
bottleneck where spatial dimensions are smallest.

\begin{figure}[H]
    \centering
    \includegraphics[width=\textwidth]{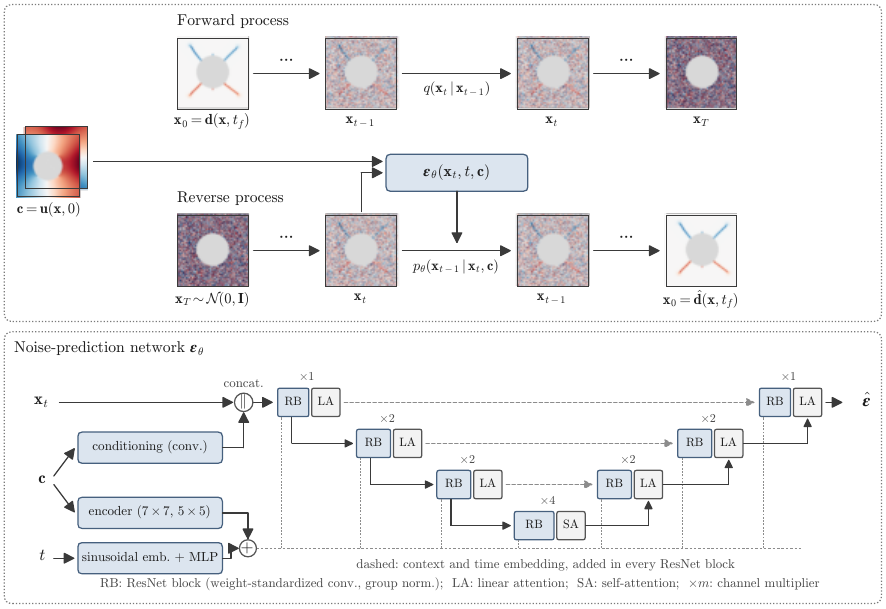}
    \caption{Schematic overview of the proposed Conditional Denoising Diffusion Probabilistic Model (cDDPM) framework. The top row illustrates the Forward Process, where the ground truth phase-field data is gradually corrupted into Gaussian noise over discrete timesteps. The bottom row depicts the Reverse Process, where the model iteratively denoises the latent representation to reconstruct the phase-field sequence $\mathbf{d}(\mathbf{x}, t_f) \in \mathbb{R}^2$, conditioned on the initial displacement field $u_x(t=0)$. The inset highlights the U-Net architecture used for noise prediction, featuring a symmetric encoder-decoder structure with ResNet blocks, weight standardization, and skip connections that facilitate multi-scale feature learning.}
    \label{fig:cddpm_d1}
\end{figure}

For \textbf{Dataset~1}, the model predicts the vector phase 
field $\mathbf{d}(\mathbf{x}, t_f)$ at the final load step, 
conditioned on the initial displacement field 
$\mathbf{u}(\mathbf{x}, 0)$. Conditioning is applied 
at two scales. Locally, a convolutional conditioning 
module aligns the displacement feature map with the 
noisy input and concatenates it at the first layer, 
providing pixel-wise geometric guidance. Globally, a 
separate deep encoder with $7 \times 7$ and 
$5 \times 5$ convolutional stages produces a latent 
context vector that is injected into each ResNet block 
via sinusoidal time embeddings and MLP projections. 
Channel multipliers $(1, 2, 2, 4)$ progressively 
increase feature depth while reducing spatial 
resolution. The forward chain uses $T = 500$ noise 
steps. The network is trained with the noise-matching loss of 
\Cref{eq:ddpm_loss}, and the masked MSE of \Cref{eq:loss_dataset1} 
is evaluated on the final denoised prediction. The Dataset~1 cDDPM architecture 
is illustrated in \Cref{fig:cddpm_d1}.

For \textbf{Dataset~2}, the cDDPM is applied in the same 
way, predicting the scalar phase-field $d(\mathbf{x}, t_f)$ 
at the final load step conditioned on the concatenated inputs 
$d(\mathbf{x}, 0)$ and $E(\mathbf{x})$. The forward chain 
corrupts the target $d(\mathbf{x}, t_f)$ over $T = 500$ 
steps, and the reverse chain denoises conditioned on the 
initial configuration and material property field. The same 
UNet backbone, cosine noise schedule, weight-standardized 
convolutions, and $\ell_1$ noise-matching loss of 
\Cref{eq:ddpm_loss} are used for training, and the MSE of 
\Cref{eq:loss_dataset2} is evaluated on the final denoised 
prediction.

\subsection{Training and evaluation protocols}
\label{sec:training}
To make the architectures comparable, the dataset partitions, the input fields, the test metric, and the hardware are the same for all of them, and all experiments are conducted on a single NVIDIA A100 GPU. The optimizer settings and the training loss follow each architecture's formulation. For both datasets, input variables are normalized using min--max scaling computed exclusively from the training set. Given a field $v$, the normalized variable is defined as
$
\tilde v = 2(v-v_{\min})/(v_{\max}-v_{\min})-1,
$
where $v_{\min}$ and $v_{\max}$ are determined from the training data and subsequently applied unchanged to the test set. For Dataset~1, the displacement components $u_1(\mathbf{x},0)$ and $u_2(\mathbf{x},0)$ are normalized independently. For Dataset~2, only the Young's modulus field $E(\mathbf{x})$ is normalized, while the initial phase-field $d(\mathbf{x},0)\in[0,1]$ is used directly. The training losses are summarized in \Cref{tab:training_losses}. For Dataset~1, FNO, DeepONet, and UNet are trained with the masked MSE of \Cref{eq:loss_dataset1}, in which the hole region $\Omega_{\mathrm{hole}}$ is excluded. For Dataset~2, UNet is trained with the MSE of \Cref{eq:loss_dataset2}, whereas FNO and DeepONet are trained with a weighted mean absolute error (MAE) that compensates for the severe imbalance between damaged and undamaged regions,
\begin{equation}
\label{eq:wmae_dataset2}
    \mathcal{L}_{\text{wMAE}} = \frac{1}{N_\Omega} 
    \sum_{\mathbf{x} \in \Omega} w(\mathbf{x})\,
    \bigl|d_{\text{pred}}(\mathbf{x}, t_f) 
    - d_{\text{true}}(\mathbf{x}, t_f)\bigr|,
    \qquad
    w(\mathbf{x}) = 
    \begin{cases}
        6 & \text{if } d_{\text{true}}(\mathbf{x}, t_f) > 0.5, \\
        1 & \text{otherwise.}
    \end{cases}
\end{equation}
Two architectures cannot use these field-space losses by construction. Latent DeepONet is trained on the latent codes produced by its pretrained autoencoder, with an MSE for Dataset~1 and an MAE for Dataset~2, and the cDDPM is trained with the noise-matching loss of \Cref{eq:ddpm_loss}. Because the training losses differ, every comparison in \Cref{sec:results} uses a single test metric per dataset, evaluated on the predicted field in physical space: the masked MSE of \Cref{eq:loss_dataset1} for Dataset~1 and the MSE of \Cref{eq:loss_dataset2} for Dataset~2.

\begin{table}[H]
\small
\centering
\renewcommand\arraystretch{1.0}
\caption{Training loss used for each architecture. All architectures 
are scored on the test set with the same metric, regardless of the 
loss used for training. ``Weighted MAE'' refers to 
\Cref{eq:wmae_dataset2}. Latent DeepONet is trained on the latent 
codes of a pretrained autoencoder, which is itself trained with an 
MSE reconstruction loss.}
\label{tab:training_losses}
\begin{tabular}{lll}
    \toprule
    \textbf{Architecture} & \textbf{Dataset~1} & \textbf{Dataset~2} \\
    \midrule
    FNO             & Masked MSE & Weighted MAE \\
    DeepONet        & Masked MSE & Weighted MAE \\
    Latent DeepONet & MSE on latent codes & MAE on latent codes \\
    UNet            & Masked MSE & MSE \\
    cDDPM           & $\ell_1$ noise-matching loss & $\ell_1$ noise-matching loss \\
    \midrule
    Test metric (all) & Masked MSE, \Cref{eq:loss_dataset1} & MSE, \Cref{eq:loss_dataset2} \\
    \bottomrule
\end{tabular}
\end{table}

For Dataset~1, training is performed in two stages. First, each model is optimized until the test error satisfies $\mathrm{MSE}<6\times10^{-5}$; the corresponding wall-clock time is recorded as a measure of training efficiency. Training then continues from the same checkpoint until either convergence is observed or a maximum training budget of 24 hours is reached, yielding the minimum attainable test error for the architecture. For Dataset~2, each model is trained for a fixed number of epochs, long enough for the test error to plateau (\Cref{fig:mse_plots_dataset2}), and the checkpoint with the lowest test error is retained; no separate validation set is used. Performance is quantified using four metrics: (i) prediction accuracy, measured by the masked MSE (\Cref{eq:loss_dataset1}) for Dataset~1 and the MSE (\Cref{eq:loss_dataset2}) for Dataset~2 on the held-out test set; (ii) training efficiency, measured by the time required to attain the threshold $\mathrm{MSE}<6\times10^{-5}$ for Dataset~1 and the total time to convergence for Dataset~2; (iii) inference cost, measured as the average wall-clock time per test sample following a warm-up pass; and (iv) model complexity, measured by the total number of trainable parameters. To assess temporal super-resolution on Dataset~1, an additional interpolation experiment is conducted. Each surrogate is trained using Case~2, which contains $N_t=22$ snapshots obtained by retaining every second state of the full 44-step trajectory, and is subsequently evaluated on the complete 44-step sequence. Let $\mathcal{T}_{\mathrm{obs}}$ denote the set of training time steps and $\mathcal{T}_{\mathrm{int}}$ the complementary set of withheld intermediate steps. Prediction errors are reported over $\mathcal{T}_{\mathrm{obs}}$ and over the full sequence $\mathcal{T}_{\mathrm{obs}} \cup \mathcal{T}_{\mathrm{int}}$, so that the increase between the two measures the cost of interpolating to unseen temporal locations.

\subsection{Using and extending Fracture-Bench}
\label{sec:extending}

The benchmark tasks of \Cref{sec:problem_statement} are one way of using the datasets, chosen so that all five architectures could be compared on equal terms. They are not the only way.

\paragraph{Other tasks on the same data.} Both datasets store the displacement and phase-field solutions at every load step, so they support tasks beyond those benchmarked here. Examples include predicting the full 31-step crack evolution in Dataset~2 rather than only its final state, predicting the displacement field or both fields jointly, step-by-step (autoregressive) prediction of crack growth, and extrapolation tests in which one inclusion configuration is withheld from training. Because the governing equations are known, the data can also be combined with physics-informed training objectives. The loss function, the preprocessing, and the training budget are the user's choice.

\paragraph{Adding an architecture.} To make a new surrogate comparable with the reference results of \Cref{sec:results}, it should be trained on the training split and evaluated on the test split of \Cref{tab:regime}, and the following should be reported: the test error in the metric of \Cref{eq:loss_dataset1} for Dataset~1 or \Cref{eq:loss_dataset2} for Dataset~2, the training wall-clock time and hardware, the inference time per sample, and the number of trainable parameters.

\paragraph{Adding a fracture problem.} A new dataset can be added to the benchmark if it provides (i) the solution fields on a uniform grid in the same array layout as the existing datasets (samples $\times$ load steps $\times$ grid points $\times$ field components), (ii) the simulation code or input files needed to regenerate it, (iii) a fixed train--test split, and (iv) a statement of the learning task, namely its inputs, its outputs, and the metric used to score it. A baseline result from at least one existing architecture is encouraged. Contributions can be submitted as an issue or pull request in the project repository, \url{https://github.com/Centrum-IntelliPhysics/fracture_bench}, or by contacting the authors.

\paragraph{Interactive dashboard.} A companion website, \url{https://fracture-bench.github.io}, lets users explore both datasets and the trained surrogates without writing code. For each dataset, the user selects one of ten held-out test simulations and compares the predictions of FNO, DeepONet and Latent DeepONet with the finite element solution and with pixel-wise error maps. For Dataset~1, the predicted crack evolution can be stepped through at 11, 22 or 44 load steps.

\section{Results and discussion}
\label{sec:results}

This section presents a systematic evaluation of the five surrogate 
architectures - FNO, DeepONet, Latent DeepONet, UNet, and diffusion 
models - on both datasets. For each dataset, results are organized 
across four studies: prediction accuracy, computational efficiency, 
temporal generalization, and qualitative assessment of predicted 
crack fields. A final subsection compares neural operator predictions 
against linear temporal interpolation as a non-learning baseline.

\subsection{Dataset 1: Square plate with a central hole}
\label{sec:results_d1}

The aggregate performance metrics are summarized in \Cref{tab:computational_profile_combined}. Across all temporal resolutions, FNO consistently achieves the best combination of accuracy and training efficiency, reaching the prescribed error threshold substantially faster than the remaining architectures while also attaining the lowest final test error. This behavior is consistent with the spectral parameterization of the operator kernel, which efficiently captures the global spatial structure governing crack evolution.

Latent DeepONet provides the strongest performance among the remaining architectures. Relative to DeepONet, the latent formulation converges considerably faster and achieves substantially lower error, indicating that learning the temporal operator in a compressed latent manifold is more effective than operating directly in the high-dimensional physical space. Although its final accuracy remains slightly below that of FNO, the gap is small compared to the substantial reduction in model complexity.

UNet exhibits stable convergence across all temporal resolutions and achieves competitive accuracy. Its performance improves as the number of temporal snapshots increases, which is attributed to the three-dimensional encoder--decoder architecture. For short sequences, repeated downsampling in the temporal direction introduces padding artifacts that become progressively less influential as the temporal extent grows.

In contrast, DeepONet scales poorly with temporal resolution. Both optimization difficulty and final error increase markedly with sequence length, and the model fails to reach the target accuracy for the most demanding case. The cDDPM exhibits a similar trend. While successful for the coarsest temporal resolution, its performance deteriorates as the temporal dimension increases, suggesting that conditioning solely on the final crack state provides an insufficient training signal for learning the underlying spatio-temporal evolution.

The computational trade-offs are also evident from \Cref{tab:computational_profile_combined}. FNO offers the lowest training cost but incurs relatively expensive inference due to repeated Fourier transforms. DeepONet occupies the opposite extreme, requiring the longest training time while providing the fastest inference. Latent DeepONet and UNet occupy an intermediate regime, balancing training efficiency, prediction accuracy, and inference cost. The cDDPM is the most expensive architecture at inference because each prediction requires execution of the full reverse diffusion chain.

\begin{table}[H]
    \centering
    \small
    \renewcommand\arraystretch{1.0}
    \caption{Consolidated computational profile across all 
    temporal resolutions for Dataset~1. Training time is 
    measured to reach a test MSE of $6 \times 10^{-5}$. 
    Minimum test MSE is the lowest value achieved after 
    extended training under the 24-hour wall-clock budget, 
    averaged across all test samples and time steps. }
    \label{tab:computational_profile_combined}
    \begin{tabular}{llccccc}
        \toprule
        \textbf{Metric} & \textbf{Steps} & \textbf{FNO} 
        & \textbf{DeepONet} & \textbf{Latent DON} 
        & \textbf{UNet} & \textbf{cDDPM} \\
        \midrule
        \multirow{3}{*}{Training time (hr)}
            & 11 & 0.1 & 5   & 3 & 1.5 & 6.5 \\
            & 22 & 0.2 & 20  & 3 & 2.1 & \nd \\
            & 44 & 0.3 & \nd & 4 & 3.5 & \nd \\
        \midrule
        \multirow{3}{*}{Inference time (s)}
            & 11 & $4{\times}10^{-1}$   & $1{\times}10^{-3}$  
                & $6{\times}10^{-3}$  & $17.7{\times}10^{-3}$ 
                & 13.42 \\
            & 22 & $4{\times}10^{-1}$   & $1{\times}10^{-3}$  
                & $6{\times}10^{-3}$  & $34.8{\times}10^{-3}$ 
                & \nd   \\
            & 44 & $4.5{\times}10^{-1}$ & \nd                 
                & $10{\times}10^{-3}$ & $69.8{\times}10^{-3}$ 
                & \nd   \\
        \midrule
        \multirow{3}{*}{Parameters ($\times 10^6$)}
            & 11 & 314 & 3   & 48 & 98 & 22.4 \\
            & 22 & 314 & 5   & 61 & 98 & \nd  \\
            & 44 & 314 & \nd & 87 & 98 & \nd  \\
        \midrule
        \multirow{3}{*}{Min. test MSE ($\times 10^{-6}$)}
            & 11 & 1 & 29  & 0.9 & 20 & \nd \\
            & 22 & 1 & 50  & 3   & 9  & \nd \\
            & 44 & 2 & \nd & 7   & 10 & \nd \\
        \bottomrule
    \end{tabular}
\end{table}

\begin{figure}[H]
    \centering
    \includegraphics[width=\textwidth]{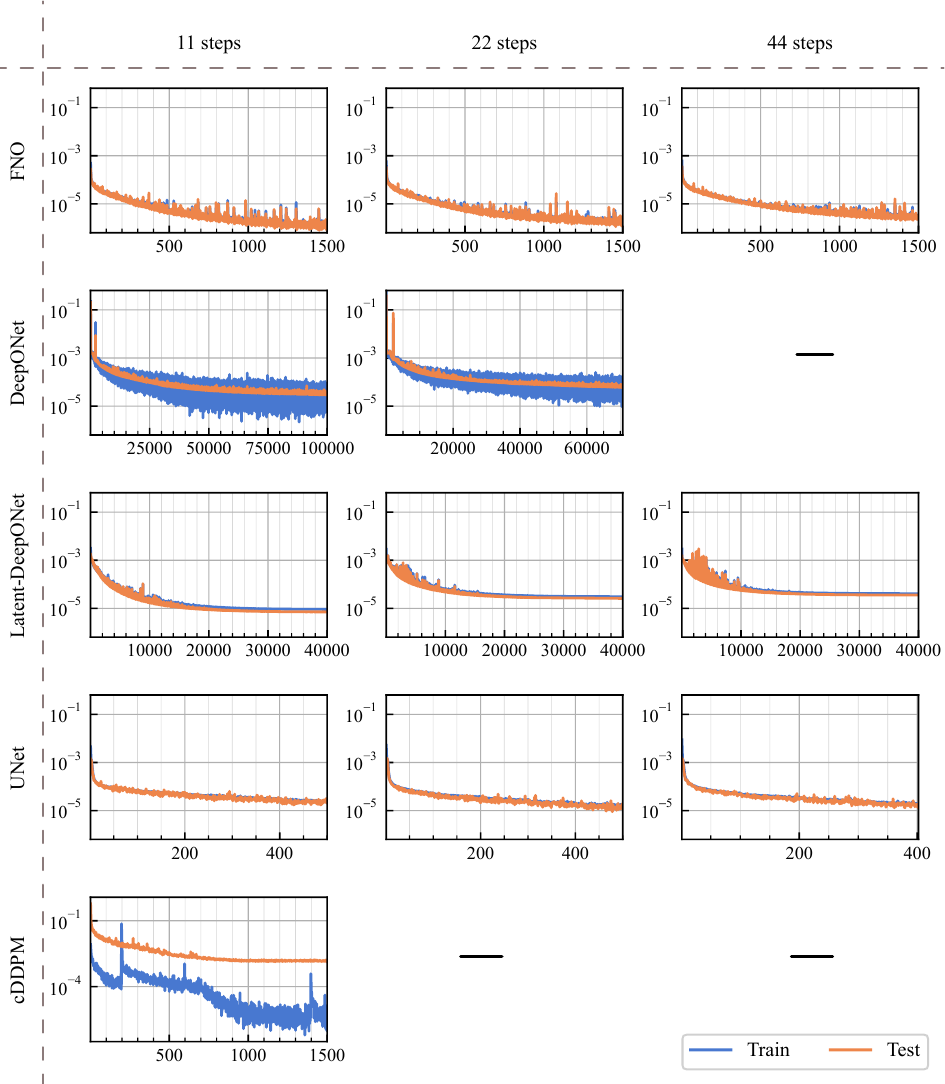}
\caption{Training (blue) and test (orange) MSE versus 
    training epoch for each surrogate architecture and 
    each temporal resolution ($N_t \in \{11, 22, 44\}$ load 
    steps) in Dataset~1. Each row corresponds to one 
    architecture and each column to one temporal case. 
    FNO and UNet converge within hundreds to a few 
    thousand epochs; DeepONet requires up to $10^5$ epochs 
    to approach the same MSE level and does not converge at 
    $N_t = 44$ steps within the 24-hour budget (row left 
    blank); cDDPM converges at 11 steps but plateaus well 
    above the threshold at 22, and 44 steps. The Latent 
    DeepONet curves are evaluated in the encoded latent 
    space and therefore cannot be directly compared with 
    the other architectures on the same vertical axis, 
    which are evaluated in physical space.}
    \label{fig:mse_plots_dataset1}
\end{figure}

The convergence histories shown in \Cref{fig:mse_plots_dataset1} are consistent with the aggregate statistics of \Cref{tab:computational_profile_combined}. FNO and UNet exhibit rapid and stable optimization, whereas DeepONet requires substantially more training iterations to reach comparable error levels. The cDDPM converges only for the coarsest temporal discretization and plateaus at higher resolutions. Collectively, these results indicate that architectures designed to exploit spatial structure directly through convolutional or spectral representations are considerably easier to optimize than operator-learning approaches acting in the full physical space.

\begin{figure}[H]
    \centering
    \includegraphics[width=0.9\textwidth]{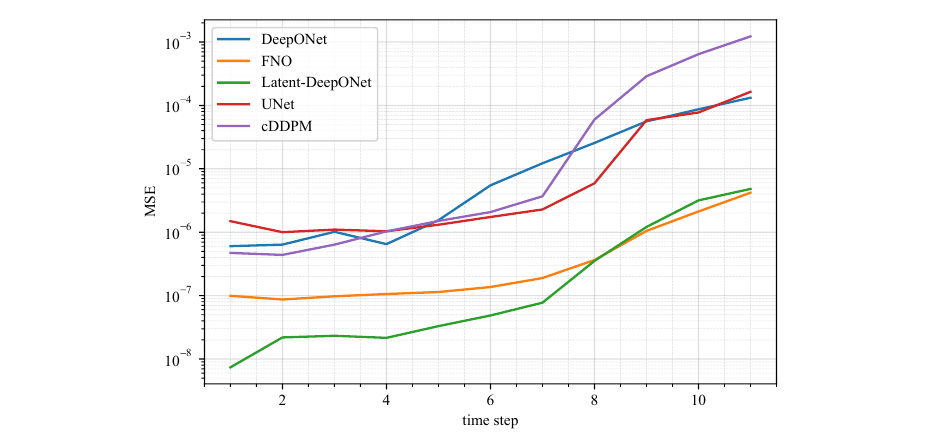}
    \caption{Test MSE per time step for the 11-step case 
    of Dataset~1. The error increases monotonically with 
    time step for all architectures as the phase-field 
    evolves toward sharp localized crack bands. Latent 
    DeepONet achieves the lowest error at early steps; 
    FNO and Latent DeepONet converge in error beyond 
    step 8; cDDPM degrades the fastest beyond step 8.}
    \label{fig:temporal_mse}
\end{figure}

The temporal error evolution for the 11-step case is shown in \Cref{fig:temporal_mse}. All architectures exhibit increasing error with load step, reflecting the transition from smooth damage fields to highly localized crack patterns. Differences among models are modest during the early stages of crack growth, but become increasingly pronounced as localization develops. FNO and Latent DeepONet remain the most accurate throughout the loading history, whereas DeepONet, UNet, and particularly the cDDPM experience a more rapid degradation near the final stages of fracture. This behavior suggests that accurately resolving the strongly nonlinear crack-tip dynamics, rather than the early diffuse damage evolution, is the primary challenge separating the surrogate architectures.

\begin{figure}[H]
    \centering
    \includegraphics[width=\textwidth]{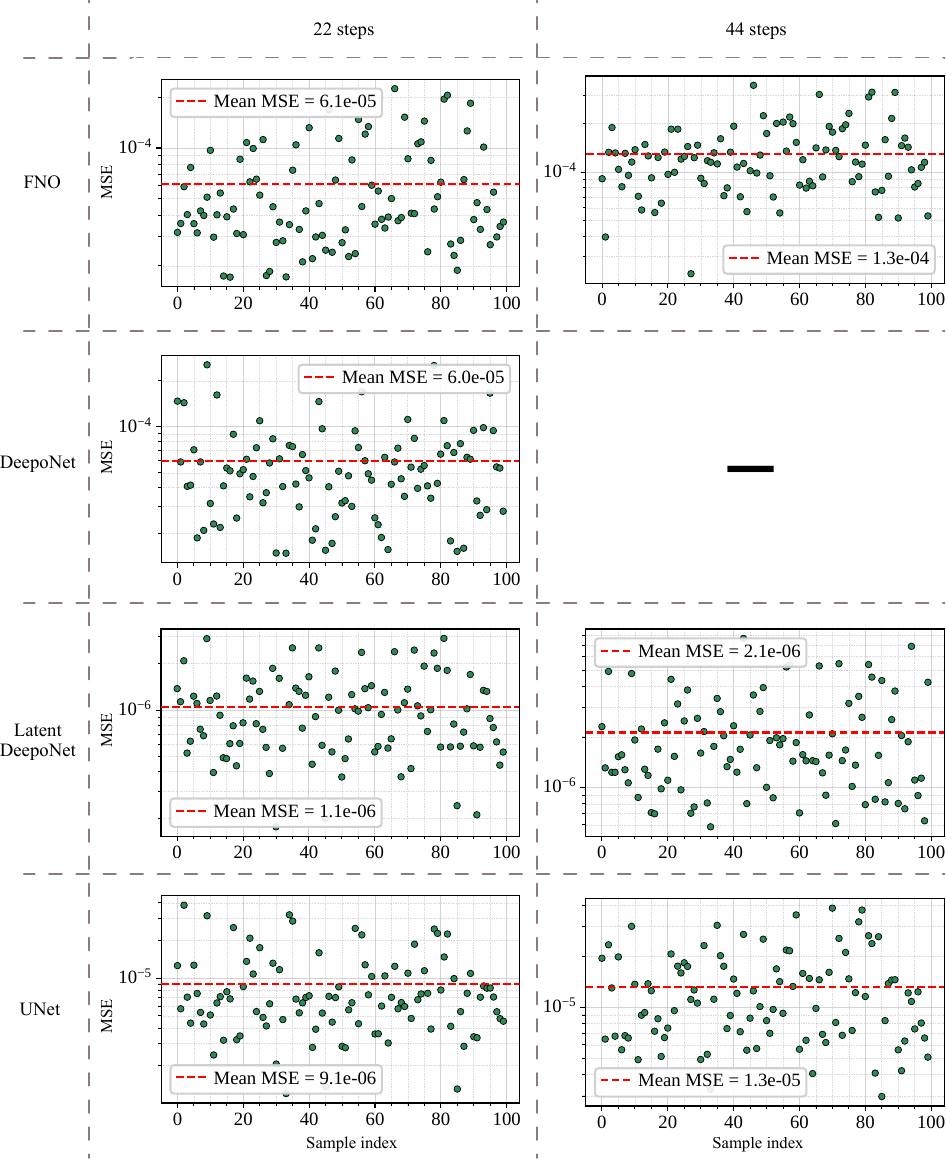}
    \caption{Per-sample test MSE for models trained on 22 load steps 
    and evaluated at both 22 steps (in-distribution) and 44 steps 
    (out-of-distribution interpolation). Each point corresponds to one test sample. The mean MSE (dashed red line) increases for all 
    architectures when evaluated at unseen intermediate steps, with 
    Latent DeepONet showing the smallest relative degradation, from 
    $1.1 \times 10^{-6}$ (22 steps) to $2.1 \times 10^{-6}$ 
    (44 steps), and FNO showing a larger increase from 
    $6.1 \times 10^{-5}$ to $1.3 \times 10^{-4}$.}
    \label{fig:interpolating_mse_per_sample}
\end{figure}

Temporal generalization is examined in \Cref{fig:interpolating_mse_per_sample}, which reports the per-sample test error of models trained on the 22-step sequence (Case~2) when they are evaluated on those 22 steps and on the full 44-step sequence, half of which was withheld during training. For both FNO and Latent DeepONet the mean error over the full 44-step sequence is roughly twice the error over the 22 training steps, from $6.1\times10^{-5}$ to $1.3\times10^{-4}$ for FNO and from $1.1\times10^{-6}$ to $2.1\times10^{-6}$ for Latent DeepONet, and the spread across test samples remains similar. Both models therefore interpolate in time without a breakdown in accuracy, with Latent DeepONet retaining the lower error.

\begin{figure}[H]
    \centering
    \includegraphics[width=\textwidth]{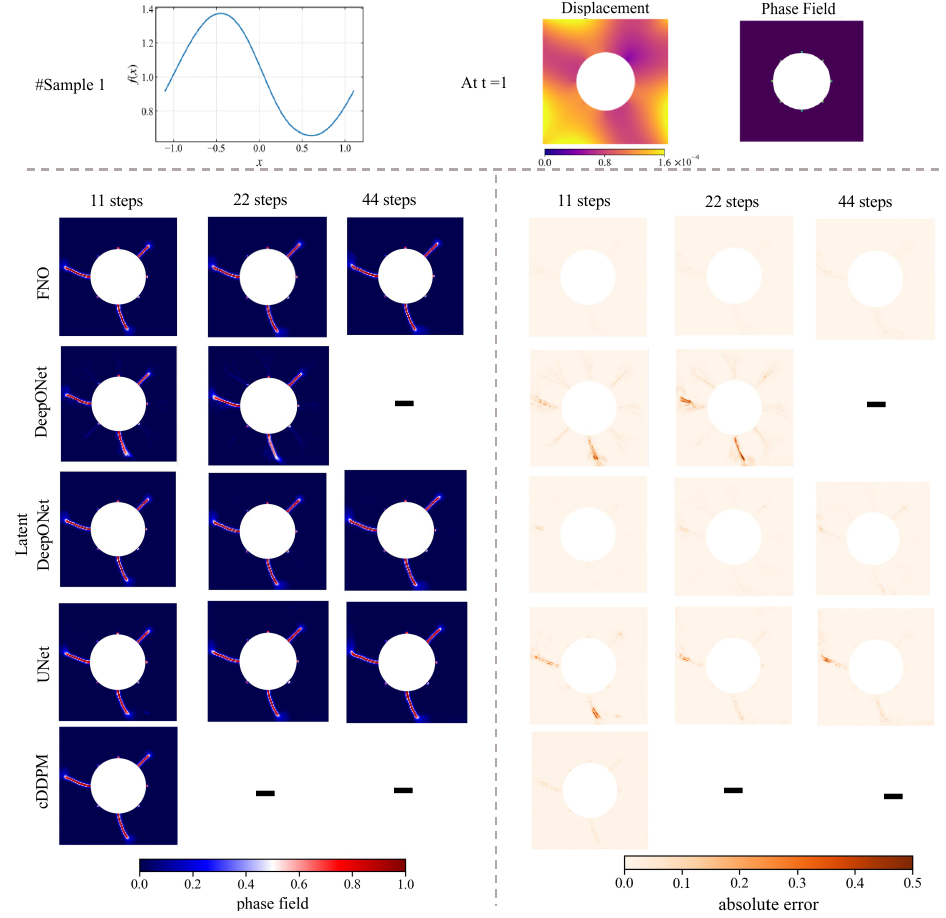}
\caption{Predicted phase-field magnitude 
    $\|\mathbf{d}(\mathbf{x}, t_f)\|$ and absolute 
    pointwise error at the final load step for 
    representative test sample~1 of Dataset~1, across 
    all five surrogate architectures and three temporal 
    resolutions ($N_t \in \{11, 22, 44\}$ load steps). 
    \textit{Top row:} GRF boundary profile $f(x)$ and 
    reference displacement magnitude and phase-field 
    at the initial load step $t = 1$. 
    \textit{Left block of each pair:} Predicted phase 
    field at the final load step $t_f$. 
    \textit{Right block of each pair:} Absolute 
    pointwise error $|\mathbf{d}_{\mathrm{pred}} - 
    \mathbf{d}_{\mathrm{true}}|$ between prediction 
    and ground truth. Black bars indicate cases where 
    the model did not converge to the threshold MSE 
    of $6 \times 10^{-5}$ within the 24-hour training 
    budget. FNO and Latent DeepONet reproduce crack 
    paths accurately at all temporal resolutions; 
    UNet produces wider crack bands at $N_t = 11$ due 
        to spatial padding in the 3D encoder; cDDPM converges 
    only at $N_t = 11$, and DeepONet at $N_t = 11$ and $22$.}
    \label{fig:sample_1_summary}
\end{figure}

Qualitative predictions for a representative test sample are shown in \Cref{fig:sample_1_summary}, which compares the predicted phase-field at the last load step of each case with the finite element reference. Every architecture that reached the error threshold recovers the correct set of cracks growing from the hole. The differences are confined to the crack bands: the errors of FNO and Latent DeepONet are barely visible at all three temporal resolutions, UNet produces slightly wider crack bands at $N_t = 11$, and DeepONet and the cDDPM show the largest errors, concentrated along the cracks. This is consistent with \Cref{fig:temporal_mse}, where the error of every model grows in the last load steps as the cracks localize.

\subsection{Dataset 2: Heterogeneous plate with two circular inclusions}
\label{sec:results_d2}

\begin{figure}[H]
    \centering
    \includegraphics[width=\textwidth]{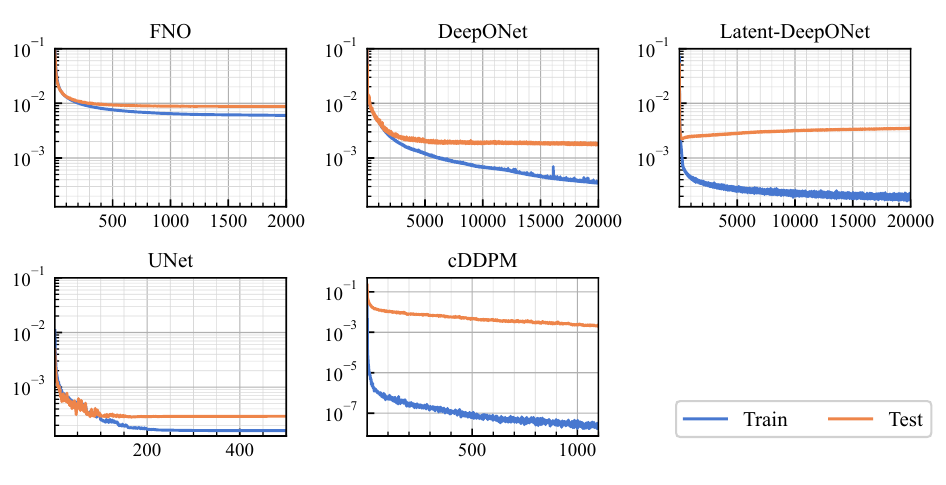}
\caption{Training (blue) and test (orange) MSE versus 
    training epoch for each surrogate architecture on 
    Dataset~2 (heterogeneous plate with circular inclusions). All models are trained jointly on all 
    four inclusion configurations. UNet converges fastest 
    in approximately 400 epochs; FNO converges within 
    2,000 epochs; DeepONet and Latent DeepONet require 
    up to 20,000 epochs; and cDDPM, while achieving the 
    lowest final test MSE, requires the longest training 
    time of 23 hours before loss plateaus.}
    \label{fig:mse_plots_dataset2}
\end{figure}

The aggregate performance metrics for Dataset~2 are summarized in \Cref{tab:computational_profile_dataset2}, while the convergence histories are shown in \Cref{fig:mse_plots_dataset2}. In contrast to Dataset~1, the ranking of the architectures changes substantially. The cDDPM achieves the lowest test error, followed by UNet and FNO, whereas DeepONet remains the least accurate model. This reversal highlights an important distinction between the two benchmarks. Dataset~1 requires learning a spatio-temporal evolution operator, whereas Dataset~2 involves only a single spatial prediction. Without the burden of modeling temporal dynamics, the diffusion model can devote its representational capacity to resolving the highly localized crack band and its interaction with material interfaces, which appears particularly well suited to iterative denoising.

UNet benefits from the same simplification. Unlike the three-dimensional architecture used in Dataset~1, the two-dimensional UNet operates on a spatial grid whose dimensions are compatible with the encoder--decoder hierarchy, eliminating the padding artifacts previously observed at coarse temporal resolutions. FNO remains competitive but is consistently less accurate than UNet and the cDDPM, suggesting that the global spectral representation is less effective at resolving localized crack-tip behavior and sharp inclusion-boundary interactions. DeepONet again exhibits the weakest performance, indicating that compressing the input fields into a fixed-dimensional latent representation is insufficient for preserving the fine spatial information required to accurately predict crack deflection around inclusions. Latent DeepONet alleviates this limitation but does not fully eliminate it.

The convergence curves in \Cref{fig:mse_plots_dataset2} reinforce these observations. UNet converges most rapidly, while the cDDPM requires substantially longer optimization but ultimately reaches the lowest error. The operator-learning models converge more slowly and plateau at higher error levels, indicating that the dominant challenge in Dataset~2 is spatial localization rather than global field reconstruction.

The computational trade-offs differ from those observed in Dataset~1. UNet provides the most favorable balance between accuracy and computational cost, achieving near-best accuracy with the shortest training time and modest inference cost. FNO offers similar efficiency but at slightly lower accuracy. Although the cDDPM delivers the most accurate predictions, its iterative reverse-diffusion process makes it several orders of magnitude more expensive at inference, limiting its practicality in applications requiring large numbers of forward evaluations. DeepONet and Latent DeepONet retain extremely low inference costs but sacrifice predictive accuracy.

\begin{figure}[H]
    \centering
    \includegraphics[width=
0.93\textwidth]{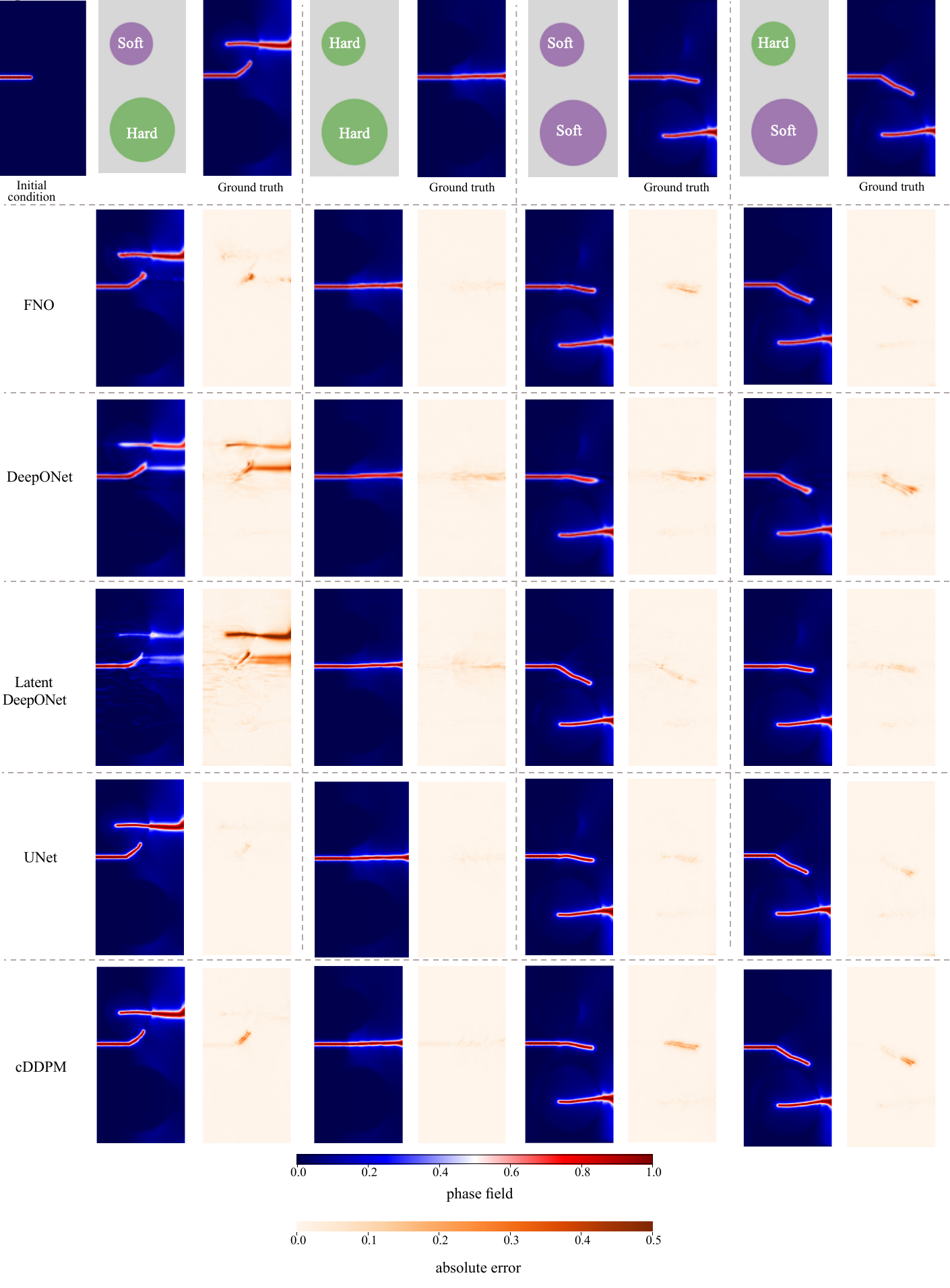}
    \caption{Ground truth phase-field 
    $d(\mathbf{x}, t_f)$ and absolute prediction errors 
    from FNO, DeepONet, Latent DeepONet, UNet, and cDDPM for five 
    Dataset~2 test samples sharing an identical initial 
    crack location but differing in inclusion 
    configuration (less stiff,\,stiffer), (stiffer,\,stiffer), 
    (less stiff,\,less stiff), and (stiffer,\,less stiff). \textit{Top row:} 
    Initial crack condition and ground truth final phase 
    field. \textit{Subsequent rows:} Predicted phase 
    field (left of each pair) and absolute error (right) 
    for each architecture. The variation in crack path 
    across the four configurations demonstrates the 
    sensitivity of fracture trajectories to inclusion 
    stiffness, and the variation in error patterns across 
    architectures reveals the differing capabilities of 
    each model in resolving the crack-tip region and the 
    inclusion-boundary transition zone.}
    \label{fig:dataset2_qualitative}
\end{figure}

Qualitative predictions for representative test cases are shown in \Cref{fig:dataset2_qualitative}. All architectures correctly reproduce the primary physical mechanism: cracks tend to deflect around stiffer inclusions and propagate through less stiff inclusions. The principal differences lie in the resolution of the crack-tip region and the inclusion-transition zone. The cDDPM and UNet generate the sharpest crack representations, with errors largely confined to the crack band itself. FNO produces comparable crack trajectories but exhibits slightly larger localization errors near material interfaces. DeepONet generates noticeably more diffuse damage fields, particularly in configurations exhibiting strong crack deflection, consistent with the limitations of its fixed-dimensional latent representation. Latent DeepONet partially mitigates this issue and generally captures the correct crack trajectory, although its accuracy degrades as the stiffness contrast increases and the crack path becomes more geometrically complex.

Overall, Dataset~2 highlights a different regime from Dataset~1. When the task is dominated by resolving a single highly localized spatial field, convolutional generative models outperform operator-learning approaches. The cDDPM provides the highest accuracy, UNet offers the best balance between accuracy and computational efficiency, and DeepONet-based models remain limited by the information bottleneck imposed by their latent representations.

\begin{table}[H]
    \centering
    \small
    \renewcommand\arraystretch{1.15}
    \caption{Computational profile for Dataset~2. All models 
    are trained jointly on all four inclusion configurations in 
    a single run. Training time is the total wall-clock time to 
    convergence on a single NVIDIA A100 GPU. Inference time is 
    per sample, averaged over the 400-sample test set. Mean 
    test MSE is averaged across all test samples on the full 
    $128 \times 256$ domain. 
    \textit{Note:} For Latent DeepONet, per-case MSE 
    values are reported as (less stiff-stiffer, stiffer-stiffer, less stiff-less stiff, 
    stiffer-less stiff).}
    \label{tab:computational_profile_dataset2}
    \newcolumntype{Y}{>{\centering\arraybackslash}X}
    \begin{tabularx}{160mm}{Y Y Y Y Y Y}
        \toprule
        \textbf{Metric}
        & \textbf{FNO} & \textbf{DeepONet}
        & \textbf{Latent DON} & \textbf{UNet} & \textbf{cDDPM} \\
        \midrule
        Training time (hr) 
            & 2.48 & 18.60 & 16.7 & 0.87 & 23.02 \\
        Inference time (s) 
            & $1.6 \times 10^{-1}$ & $6.2 \times 10^{-4}$ 
            & $4.9 \times 10^{-4}$ & $5.9 \times 10^{-3}$ & $7.06 \times 10^{2}$ \\
        Parameters ($\times 10^{6}$) 
            & 37.80 & 6.10 & 72.43 & 124.40 & 31.79 \\
        Mean test MSE ($\times 10^{-4}$) 
            & 3.85 & 17.2 & 9.57,7.82, 3.58, 4.66& 2.93 & 0.45 \\
        \bottomrule
    \end{tabularx}
\end{table}

\section{Conclusions}
\label{sec:conclusion}

We introduced Fracture-Bench, an open benchmark for machine-learning surrogates of phase-field fracture. Its main contribution is the resource itself: two datasets of 1,000 and 4,000 simulations that differ in fracture formulation, material heterogeneity and prediction task, released with the codes that generated them, fixed splits, common metrics, regime descriptors, training code and an interactive dashboard. The comparison of five architectures, FNO, DeepONet, Latent DeepONet, UNet and the cDDPM, is a first use of this resource. It provides reference results against which new methods can be measured, and it shows why a shared benchmark is needed.

The main findings are as follows.
\begin{itemize}[leftmargin=*,nosep]
    \item No architecture is best on both datasets. On the trajectory task of Dataset~1, FNO and Latent DeepONet are the most accurate (test MSE of order $10^{-6}$), and FNO trains 10 to 100 times faster than the others. On the final-state task of Dataset~2, the cDDPM is the most accurate ($0.45\times10^{-4}$), and UNet offers the best balance of accuracy (second-best error), training time (0.87~h) and inference time (6~ms).
    \item Accuracy and cost trade off strongly. The most accurate model on Dataset~2 is also the slowest by five orders of magnitude at inference, and the fastest models at inference (DeepONet and Latent DeepONet) are among the least accurate.
    \item The error of every architecture is concentrated where and when the crack localizes, and on Dataset~2 near crack-path bifurcations, where inputs less than one grid cell apart lead to different crack paths. Domain-averaged errors understate these failures, and for near-bifurcation inputs a probabilistic prediction is more appropriate than a single field.
    \item The random splits test interpolation in low-dimensional, densely sampled input spaces. Generalization to new loadings, configurations and crack locations must be measured separately, with dedicated splits, for example by withholding one inclusion configuration from training (\Cref{sec:extending}).

\end{itemize}

The tasks studied here use only part of what the datasets contain. The stored displacement fields and intermediate load steps support other learning problems, and users are free to choose their own loss functions and training strategies. Fracture-Bench is also meant to grow. \Cref{sec:extending} sets out what is needed to add a new architecture or a new fracture problem, including a regime descriptor that records what each dataset covers and what it lacks. We particularly invite datasets from the regimes that open data cover least: mixed-mode and shear loading, three-dimensional crack growth, dynamic, cyclic and ductile fracture, and simulations paired with experiments. Together, such contributions would show where surrogate predictions of fracture can be trusted and where more data are needed. Dataset~1 is available at \url{https://doi.org/10.7281/T1XFF19O} \cite{T1XFF19O_2025} and Dataset~2 at \url{https://doi.org/10.7281/T1RZFI3I} \cite{T1RZFI3I_2025}. The training and evaluation code is available at \url{https://github.com/Centrum-IntelliPhysics/fracture_bench}, and the interactive dashboard at \url{https://fracture-bench.github.io}.

\section*{Data and code availability}
Both datasets are openly available from the Johns Hopkins Research Data Repository: Dataset~1 at \url{https://doi.org/10.7281/T1XFF19O} \cite{T1XFF19O_2025} and Dataset~2 at \url{https://doi.org/10.7281/T1RZFI3I} \cite{T1RZFI3I_2025}. The training and evaluation code is available at \url{https://github.com/Centrum-IntelliPhysics/fracture_bench}. An interactive dashboard that shows the trained surrogates on held-out test cases is available at \url{https://fracture-bench.github.io}.

\bibliographystyle{elsarticle-num-names}
\bibliography{ref} 
\end{document}